\documentclass[amsmath,twocolumn,superscriptaddress,prb,aps]{revtex4}
\usepackage{amsthm,amsfonts,graphicx,verbatim, color}
\usepackage{graphicx}
\usepackage{bm}
\usepackage[utf8]{inputenc}
\let\oldmarginpar\marginpar
\renewcommand\marginpar[1]{\-\oldmarginpar[\raggedleft\footnotesize #1]%
{\raggedright\footnotesize #1}}

\newcommand{\be}{\begin{equation}}
\newcommand{\ee}{\end{equation}}
\newcommand{\beq}{\begin{equation}}
\newcommand{\eeq}{\end{equation}}
\newcommand{\bea}{\begin{eqnarray}}
\newcommand{\eea}{\end{eqnarray}}

\renewcommand{\epsilon}{\varepsilon}
\renewcommand{\vec}[1]{{\bf #1}}

\renewcommand{\cite}[1]{[\onlinecite{#1}]}

\begin{document}
\title{Superconductivity from weak repulsion in hexagonal lattice systems}
\author{Rahul Nandkishore}
 \affiliation{Princeton Center for Theoretical Science, Princeton University, Princeton, New Jersey 08544, USA}
 \affiliation{Department of Physics, Massachusetts Institute of Technology, Cambridge, MA02139, USA}
\author{Ronny Thomale}
\affiliation{Theoretische Physik I, Universit\"at W\"urzburg, am Hubland, 97074 W\"urzburg, Germany}
 \author{Andrey V. Chubukov}
 \affiliation{Department of Physics, University of Wisconsin, Madison, WI 53706, USA}
\date{\today}


\begin{abstract}
We analyze the pairing instabilities for fermions on hexagonal
lattices (both honeycomb and triangular ones) in a wide range of
fermionic densities ranging from Van Hove density at which a single
large Fermi surface splits into two disconnected Fermi pockets, to a
density at which disconnected pockets shrink to Fermi points
(half-filling for a honeycomb lattice and full filling for a
triangular lattice). We argue that for a generic doping in this range,
superconductivity at weak coupling is of Kohn-Luttinger type, and,
 due to the presence of
  electronic interactions beyond on-site repulsion, is a threshold
phenomenon, with superconductivity emerging only if the attraction
generated by the Kohn-Luttinger mechanism exceeds the bare repulsion
in some channel.
 For disconnected Fermi pockets, we predict that Kohn-Luttinger superconductivity, if it occurs, is likely to be $f$-wave. While the Kohn-Luttinger analysis is adequate over most of the doping range, a more sophisticated analysis is needed near Van Hove doping. We treat Van Hove doping using a parquet renormalization group, the equations for which we derive and analyze. Near this doping level, superconductivity is a universal phenomenon, arising from any choice of bare repulsive interactions. The strongest pairing instability is into a chiral $d-$wave state ($d+id$).  At a truly weak coupling, the strongest competitor is a spin-density-wave instability, however, $d-$wave superconductivity still wins. Moreover, the feedback of the spin density fluctuations into the Cooper channel significantly enhances the critical temperature over the estimates of the Kohn Luttinger theory. We analyze renormalization group equations at stronger couplings and find that the main competitor to $d-$wave supoerconductivity away from weak coupling is actually ferromagnetism. We also discuss the effect of the edge fermions and show that they are unimportant in the asymptotic weak coupling limit, but may give rise to, e.g., a charge-density-wave order at moderate coupling strengths.
\end{abstract}
\maketitle

\section{Introduction}

Understanding the instabilities of Fermi liquids has been an enduring theme of research in condensed matter physics for decades \cite{AGD}.
All rotationally-isotropic Fermi liquids
 display an instability to superconductivity in
 some non-zero
 angular momentum channel \cite{KL,ShankarRMP,Ch_maiti}. This condition does not hold for
 non-isotropic
 Fermi systems, which may remain in the normal state down to $T=0$. On the other hand, if a lattice system has an attractive interaction in some pairing channel, it may get enhanced
  when e.g., the Fermi surface (FS) is nested.
 The energy scale for
   superconductivity
   can also be enhanced by divergences in the FS density of states
     at certain doping levels.

    Of particular interest are situations where a Fermi liquid
    instability to superconductivity is driven purely by
     an electron-electron
     interaction
      rather than interaction with phonons.
   This line of research has a long history.
Pairing due to
 direct
  fermion-fermion  interactions was discussed in connection with the superfluidity of
$^3He$~\cite{3he,3he2} and became mainstream after the discovery of high-temperature superconductivity in the cuprates \cite{bednorz,M2S} and  subsequent discovery of superconductivity in $Fe$-based pnictides\cite{bib:Hosono}.

 The weak coupling scenario of the pairing due to nominally repulsive Coulomb interaction is generally known as Kohn-Luttinger (KL) mechanism, developed in 1965
 (Refs. [\onlinecite{KL,L}]).
 The KL mechanism is based upon two fundamental earlier results.
It was discovered
 by Friedel~\cite{friedel} in the early 1950s that the screened Coulomb potential in a Fermi liquid has a long-range oscillatory tail $\cos(2k_F r + \phi_0)/r^3$ at large distances $r$,
 hence at some large enough
  $r$  the screened Coulomb interaction gets over-screened and
becomes attractive.
 Next, Landau and Pitaevskii analyzed the pairing in an isotropic Fermi liquid at
non-zero orbital momentum $l$ of the pair and found that the
pairing problem decouples between different $l$.
 Because of this decoupling, if only one
partial component of the interaction is attractive and all others are
repulsive, the system already undergoes a pairing instability into
a state with $l$ for which the interaction is attractive~\cite{LandauLifshitz}

\begin{figure}
 \includegraphics[width = \columnwidth]{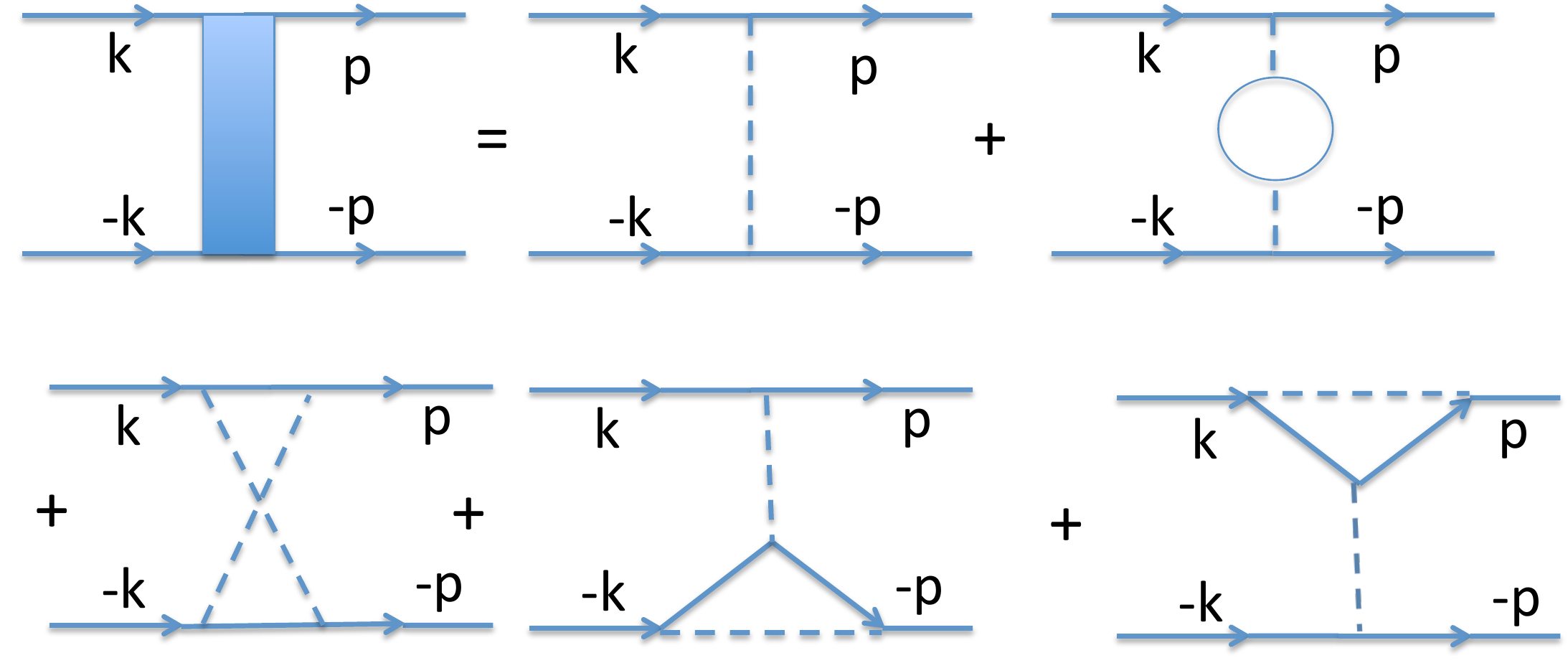}
 \caption{\label{fig:1} Kohn Luttinger mechanism of the pairing. The irreducible pairing interaction is the sum of the Coulomb interaction, which includes all regular contribution from screening ($U(q)$, represented by a dashed line), and non-analytic terms, which appear at second order in $U(q)$.
 Kohn and Luttinger demonstrated that for rotationally isotropic systems, the partial components of the irreducible pairing interaction are attractive for arbitrary $U(q)$ for large values of the angular momentum $l$, at least for odd $l$.}
 \end{figure}

 Because
the components of the interaction with large $l$ come from large
distances, it is conceivable that occasional over-screening of the
Coulomb interaction at large distances may make some of partial
interaction components with large $l$ attractive. KL analyzed
 the form of the
fully screened irreducible pairing interaction at large $l$ in
three-dimensional, rotationally isotropic systems with $k^2/(2m)$
dispersion, by separating non-analytic $2k_F$ screening and
regular screening from other  momenta. KL incorporated the
latter into the effective interaction $U({\bf q}) = U(q)$ ($q
=|{\bf q}|$) and made no assumptions about the form $U(q)$ except
that it is an analytic function of $q^2$.  The full irreducible
pairing interaction is $U(q)$ plus extra terms coming from 
 the
screening (see Fig. \ref{fig:1}). An analysis of this form was first performed for the s-wave channel in \cite{Gorkov}. KL extended the analysis to non-zero $l$.
They argued that at large $l$ contributions to partial components of
the irreducible interaction from $2k_F$ scattering scale as
$1/l^4$ due to the non-analyticity of the $2k_F$ screening (this
is the same non-analyticity which gives rise to Friedel
oscillations).  At the same time,  partial components of analytic
$U(q)$ behave at large $l$ as $e^{-l}$, i.e., are much smaller.
As a result, even though the KL contribution is second order in $U$, it overwhelms the direct first-order interaction term in channels with large enough $l$.
KL explicitly computed~\cite{KL,L} the prefactor  for $1/l^4$ term and found
  that it generally depends on the parity of $l$.
  They found that
 the interaction in channels
with odd $l$ is attractive  no matter what is the form of $U(q)$.
For the highly-screened Hubbard interaction, for which $U(0)= U(2k_F) =U$, the prefactor is attractive for both even and odd $l$.
As a result, any generic rotationally-invariant system with repulsive Coulomb interaction is unstable against pairing in channels with sufficiently large $l$.
The pairing will be into a channel which has the largest attractive component.
The situation away from the asymptotic large $l$ limit is less
  certain as analytic and non-analytic contributions to  the irreducible pairing interaction are of the same order. However, one can make progress if the bare interaction $U(q)$ is weak, by doing perturbation theory in weak interactions.
 For momentum-independent $U(q) =U$ and an isotropic system, the KL
mechanism generates attraction in all channels down to $l=1$,
with the $l=1$ channel having the strongest attraction ~\cite{Fay,kagan}. For momentum-dependent interaction, $U(q)$ itself has
components for all $l$ and whether the second-order KL contribution
can overwhelm the bare interaction  depends on the
details~\cite{Ch_maiti,alex,raghu_ch}.

The situation in lattice systems is similar but not identical to that in isotropic systems.
 Namely, there is only a discrete set of orthogonal channels imposed by a specific lattice
  symmetry.
  (For
   2D systems with $C_4$ lattice symmetry there are four one-dimensional channels $A_{1g}$, $B_{1g}$, $B_{2g}$, and $A_{2g}$, and one two-dimensional $E_g$ channel). Each channel has an infinite set of eigenfunctions, which, however, are not orthogonal to each other, i.e., the notation of a single``large l" channel no longer exists. The leading eigenfunctions in each channel can be formally associated with $s-$wave ($A_{1g}$), $p-$wave ($E_g$),  $d-$wave ($B_{1g}$ and $B_{1g}$)  etc,
    however the "higher-momentum" eigenfunctions  have the same
    lattice symmetry as a leading component  in one of the channels
    and just fall into one of orthogonal subsets.
     (For a detailed
    discussion of hexagonal lattice representations and
    its association with superconducting orders see e.g. Ref.~\onlinecite{hexa-review}. There is
     an infinite number of orthogonal linear combinations of  eigenfunctions in each subset, hence an infinite number of eigenvalues, and for superconductivity only one of eigenvalues needs to be attractive.
However, there is no generic condition  that there must be attractive channels, and, moreover,  even if some combinations of eigenfunctions are
     attractive,
    there is no condition like in the isotropic case
     at large $l$,
      that the bare interaction $U(q)$ has to be vanishingly small in one of these channels.
 All this makes the analysis of the pairing in lattice systems more involved than in the isotropic case.

 \begin{figure}
 \includegraphics[width = \columnwidth]{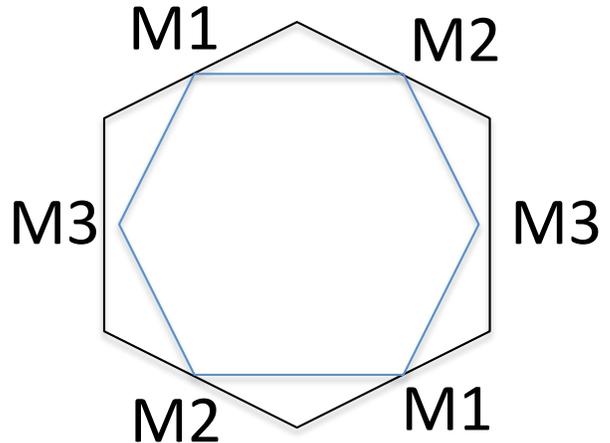}
 \caption{\label{fig:2} Fermi surface (blue lines) for fermions on a honeycomb lattice at Van-Hove doping ($\nu = 3/8$ or $5/8$, where $\nu =1/2$ corresponds to half-filling, where the fermionic spectrum has Dirac points).  For nearest-neighbor hopping, the Fermi surface consists of parallel pieces (nesting). At the end point of parallel pieces the density of states diverges (van Hove points $M_1, M_2$, and $M_3$).  Thin solid lines represent the boundaries of the Brillouin zone.}
 \end{figure}
 There are two ways to proceed and we explore both.  First, in a system with a generic FS (FS)  (i.e., the one without nesting and/or special points where the density of state diverges), one can apply perturbation theory and study KL-type superconductivity
 for a generic $U(r)$.
   For 2D systems on a tetragonal lattice, such analysis has been performed both analytically and numerically in Refs.\cite{alex,raghu_ch} (see also Ref.~\onlinecite{RonnyKagome}).
     Here we analyze KL superconductivity analytically for systems on a hexagonal lattice. We
       show
       that the subset of potential superconducting states is larger for fermions on a hexagonal lattice than on a tetragonal lattice.  We perform a KL-type calculation for a system with interaction $U(r)$ which has largest on-site (Hubbard component) but also extends to nearest and second-nearest neighbors. We show that the effective interaction taken to second order in the Hubbard $U$ gives rise to an attraction in a $d-$wave channel near Van Hove density and in a channel, which we termed as $f-$wave,
         in a range of
         dopings when the FS consists of six disconnected pockets.  This result agrees with the numerical analysis in Ref.\cite{scal_kiv_raghu}. Our study provides analytical understanding of the physics of the attraction. (We also consider pairing states not analyzed in Ref. \cite{scal_kiv_raghu}).  However, whether or not such superconductivity is actually realized depends on  how strong is the bare repulsive interaction in the corresponding channel. For Hubbard-only model, bare interaction vanishes for all non-s-wave channels, but for a generic $U(r)$ it is non-zero in all channels.
 We found that the bare interaction, taken to second neighbors, vanishes in some pairing channels, however, these channels are not the ones in which KL interaction is attractive.  As a result, KL attraction (which is second order in the bare coupling) competes with the bare repulsive interaction and must exceed it, otherwise superconductivity would not develop.
 This implies that superconductivity at a generic doping is a threshold phenomenon -- it does not occur if the interaction $U(r)$ is too weak. A somewhat richer behavior can be obtained in multilayer hexagonal lattice systems \cite{VafekSC, Guinea}, but we do not discuss these here.

 \begin{figure}
 \includegraphics[width = \columnwidth]{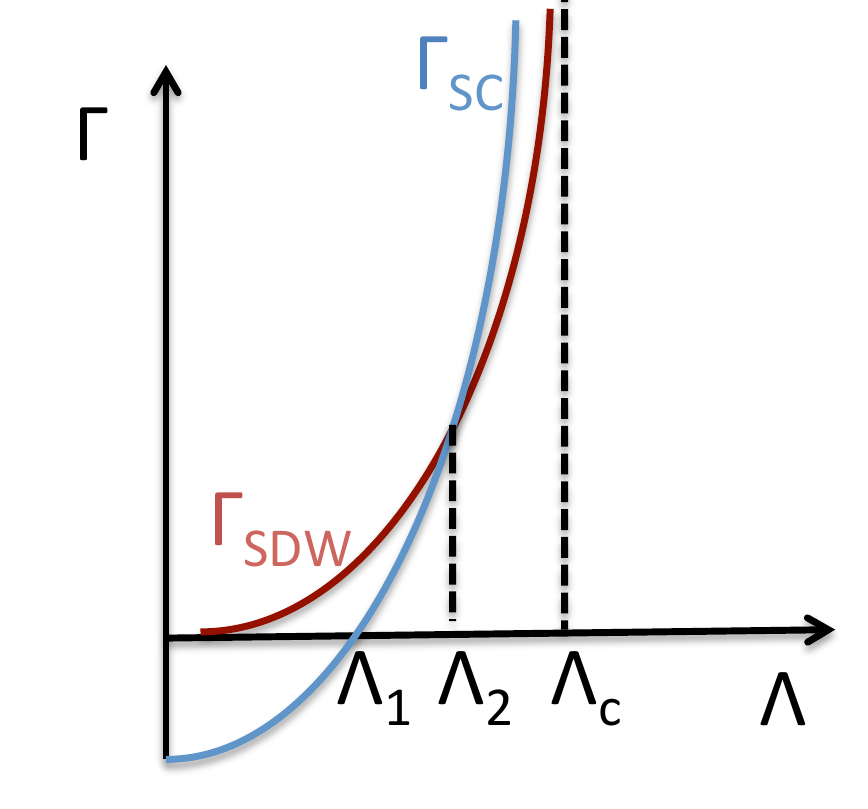}
 \caption{\label{fig:3}
 The flow of the couplings in the $s-$wave superconducting (SC) channel and spin-density-wave (SDW) channel under parquet RG at van Hove density.
  SDW coupling wins at intermediate scales. SC coupling is initially repulsive, but changes sign under RG flow and eventually wins over the coupling in SDW channel.}
 \end{figure}
 Second, at Van Hove doping, the density of states diverges at so-called Van Hove points, and, for hopping between nearest neighbors, FS consists of set of parallel lines (nesting) (see Fig.\ref{fig:2}). In this situation, KL renormalization of the bare pairing interaction is by itself logarithmically singular and has to be treated on equal footing with the pairing channel. Moreover, there are logarithmic divergences in channels corresponding to particle-hole pairing~\cite{other_channels},
  which also feed back into the particle-particle pairing channel and may generate additional attraction
   or repulsion.
    In this particular case, one can extend the analysis beyond the second order in $U(r)$. The analytical technique to do this is called parquet RG~\cite{parquetRG}.  The physics captured by parquet renormalisation group (RG) analysis  is that KL renormalization gets stronger as one progressively includes higher-order scatering events.  The parquet RG equations have been solved in the leading logarithmical approximation in Ref.\cite{ChiralSC} and the net result for superconductivity can be physically interpreted such  that the total pairing interaction (the bare one plus the KL part) evolves as the temperature gets lowered and changes from
 a repulsive one to an attractive one at some $T$ (in other words, the system self-generates the scale which is analogous to a Debye frequency).  As $T$ further decreases, the attraction grows, and at some lower $T$ the system undergoes a superconducting instability towards  a chiral $d+id$ superconducting state which breaks time-reversal symmetry. Note that at Van Hove doping, the instability is not a threshold phenomenon, but rather occurs for arbitrarily weak bare interactions.

 There is a caveat, though. In a situation when FS has Van Hove points and interactions flow as $T$ is lowered, superconductivity is not the only one option -- the system can equally well develop a density-wave instability, which in our case is a spin-density-wave instability (SDW).  It is known that superconductivity and SDW do not co-exist \cite{interplay}. At weak coupling (bare $U(r)$ is small) and exactly at Van-Hove density, superconductivity wins over SDW. What happens at larger couplings and/or slightly away from Van Hove density is less clear, and has been debated in ~\cite{Ronny1,dhlee_fawang,NandkishoreSDW, Batista, hexa-review}.
  One possibility is that SDW may win as it wins at intermediate temperature scales, and a deviation from a Van Hove density restricts the applicability of RG analysis to temperatures larger than some cutoff.    However, to properly address this issue one needs to analyze the set of RG equations beyond the leading logarithmic approximation.  This is what we will do in the second part of this paper. We show that at the Van Hove point, the physics is highly universal
   at weak coupling, and the system invariably ends up in a $d+id$ superconducting state.
   Once we move away from weak coupling, the RG approach is no longer controlled, and the neglect of higher loop diagrams can no longer be justified. However, the procedure can still be applied, although the results must be treated with caution. We show that away from weak coupling the  RG analysis that focuses purely on the hexagon corners reveals two distinct instabilities - one to d-wave superconductivity, and another to ferromagnetism. The SDW is never the leading instability if the RG is allowed to run indefinitely, but it may dominate if the RG is stopped at some intermediate energy scale by higher loop effects or self energy effects. The ferromagnetic phase is the principal new result from extending the parquet RG to strong coupling. We also discuss the effect of the edge fermions at strong coupling.  Along the ferromagnetic trajectory, the edge fermions supress ferromagnetism, and may destabilize it in the strong coupling limit towards a different phase, like a  charge-density-wave (CDW) or an s-wave superconductor. Meanwhile, along the d-wave superconducting trajectory, the edge fermions strengthen d-wave superconductivity with respect to SDW, but they may destabilise $d+id$ state at strong coupling towards another phase, such as a CDW.

 The paper is organized as follows. In the next two Sections we consider KL pairing  outside the immediate vicinity of the Van Hove density. In Sec. 2
 we consider fermions on a triangular lattice, and in Sec. 3 we consider fermions on a honeycomb lattice. In both cases we first introduce appropriate patch models and discuss potential pairing symmetries.    We then obtain  the bare pairing interaction in various channels for $U(r)$ which extends to second neighbors.
 After that, we compute second-order KL component of the pairing interaction and analyze the full pairing vertices which consist of  first-order terms and second-order KL contributions.  We argue that on both lattice a nodeless $f-$wave pairing is favorable between Van Hove density and full filling (triangular lattice) and half-filling (honeycomb lattice), and chiral $d-$wave ($d+id$) pairing is favorable at and very near Van-Hove density.   In Sec. 4 we discuss
  the system behavior {\it at} the Van Hove density. At this point the superconducting and spin density wave channels mix, so the system is described by a `parquet RG,' which we present. Our discussion follows \cite{ChiralSC}, but is substantially more detailed, and  we also extend the RG analysis to moderately strong coupling.
  In Sec. 5 we discuss the experimental situation mostly focusing on the doped graphene.  We present our conclusions in Sec. 6.

\section{Systems on a triangular lattice}
\label{sec:2}

\subsection{Fermi surface and fermionic dispersion}

\begin{figure}
 \includegraphics[width = \columnwidth]{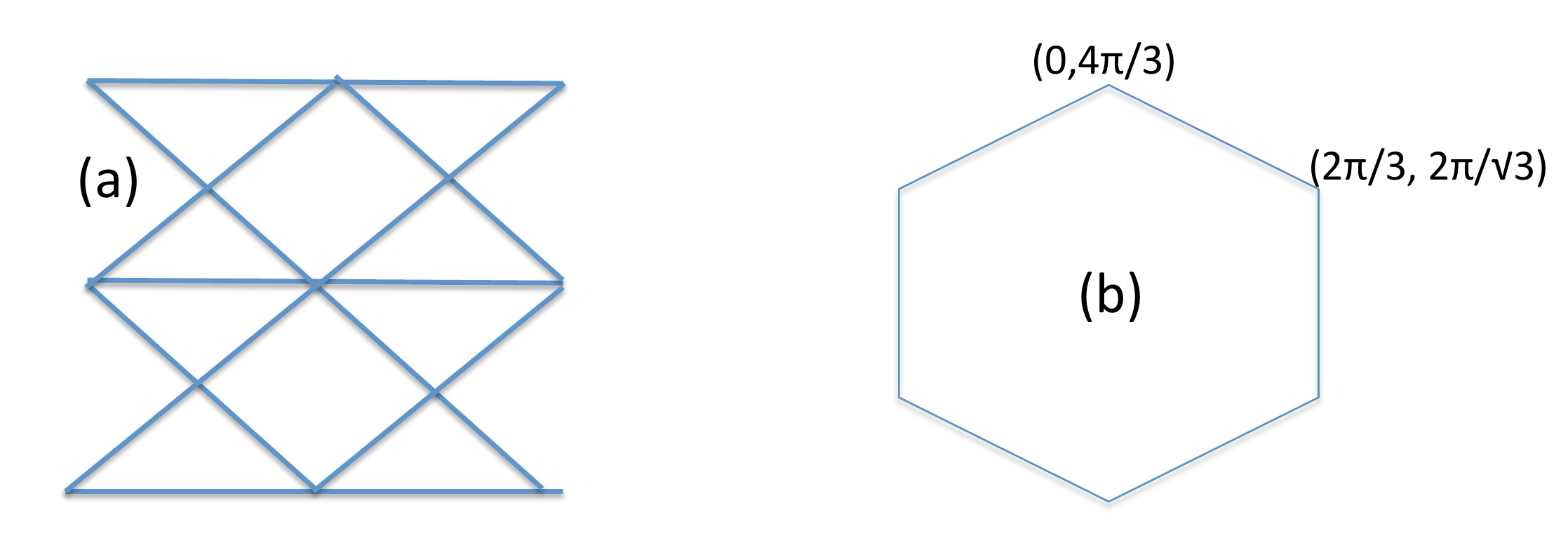}
 \caption{\label{fig:4}
 Triangular lattice (a) and the corresponding Brillouin zone in momentum space (b).}
 \end{figure}

Consider a system of fermions on a triangular lattice (Fig.\ref{fig:4}a)  with hopping $t$ between nearest neighbors. The Hamiltonian of free fermions is, in momentum space,
\beq
{\cal H}_2 = \sum_k \epsilon_k c^\dagger_k c_k
\eeq
where
\beq
\epsilon_k = -4 t \cos{\frac{k_x}{2}} \left( \cos{\frac{k_x}{2}} + \cos{\frac{k_y \sqrt{3}}{2}}\right) - \mu
\label{ch_1}
\eeq
and we set interatomic spacing $a$ to one.
The Brillouin zone (BZ) is a hexagon with corner points at $(\pm 4\pi/3,0)$ and $(\pm 2\pi/3, 2\pi/\sqrt{3})$ {\bf (Fig.\ref{fig:4}b)}

\begin{figure}
 \includegraphics[width = \columnwidth]{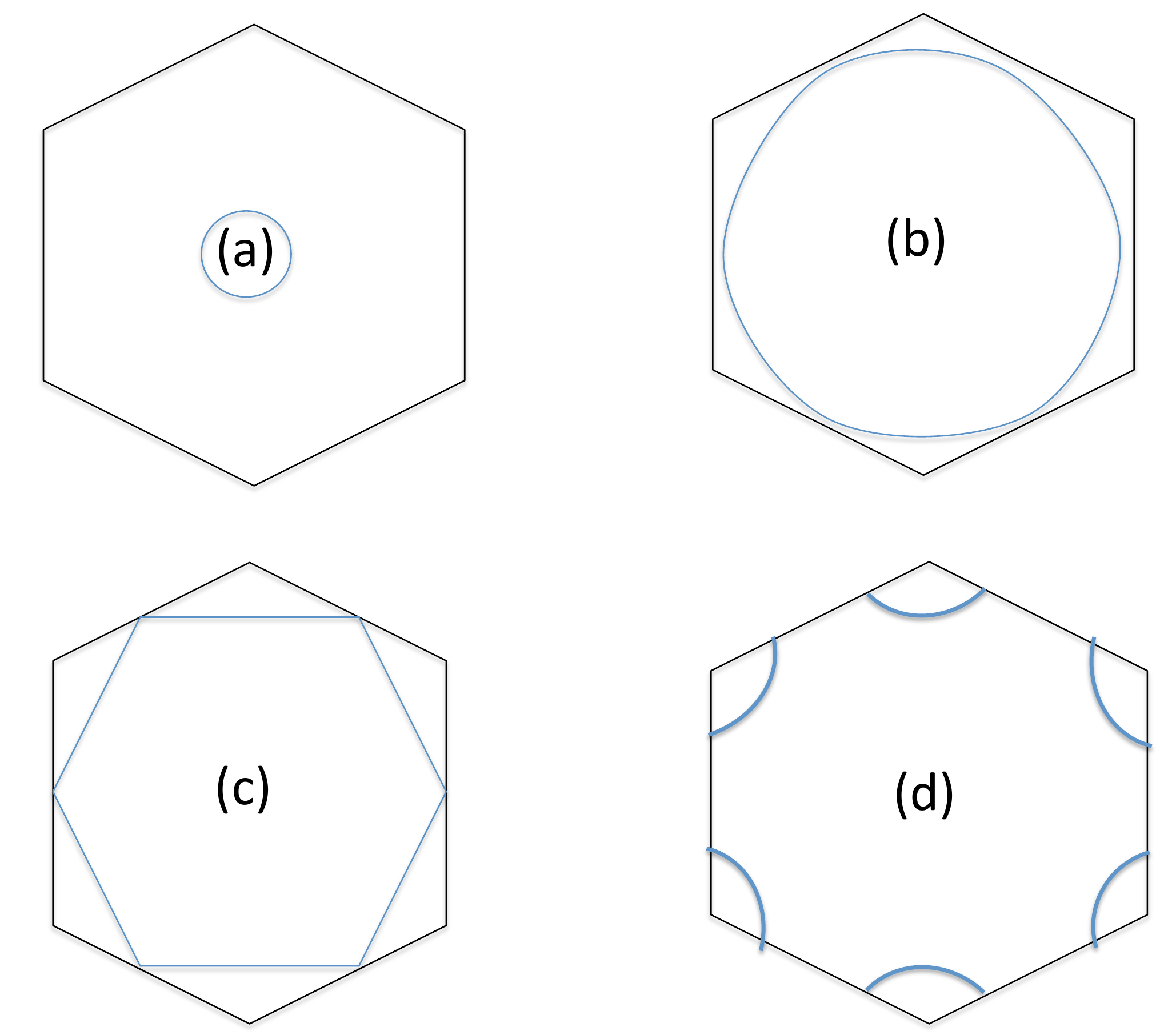}
 \caption{\label{fig:5}
 Evolution of the FS with doping for fermions on a triangular lattice.
 (a) -- $\mu$  slightly above $-8t$,  (b) --- $\mu$ approaching zero, (c) --- $\mu =0$. This is van-Hove doping. The FS consists of parallel pieces which end at van-Hove points where the density of states diverges. (d) $),\mu <t$. The FS consists of disconnected pieces.}
  \end{figure}

The topology of the FS (FS) depends on the value of the chemical potential $\mu$. At $\mu < -8t$ all states are empty at $T=0$. At $\mu =-8t+\delta$, a small, near-circular  FS opens up at the center of the BZ  (Fig.\ref{fig:5}a). As the chemical potential gets smaller, the area of the FS increases
 and its shape changes (Fig.\ref{fig:5}b). When $\mu =0$, the FS touches the BZ boundary at six van-Hove points $(0, \pm 2\pi/\sqrt{3})$ and $(\pm \pi, \pm \pi/\sqrt{3})$ (Fig.\ref{fig:5}c). Simultaneously, the FS between any nearest van-Hove points becomes a straight line, i.e., the FS contains parallel pieces (nesting) Once $\mu$ becomes positive, each van-Hove point splits into two, and the formerly singly connected FS decouples into 6 disconnected pockets centered at the corners of the BZ (Fig.\ref{fig:5}d).
  At $\mu =t$, the six FSs shrink to points at the BZ boundary at $(\pm \frac{4\pi}{3},0)$ and $(\pm \frac{2\pi}{3}, \pm \frac{3\pi}{\sqrt{3}})$.

 The van-Hove points survive even if the hopping extends beyond
 nearest neighbors.  However, a flat FS acquires a finite curvature
 once the hopping between second neighbors is added (see also Ref.~\onlinecite{hexa-review}).

\subsection{The pairing interaction}

Throughout this paper we assume that the bare interaction between low-energy fermions is some short-range interaction
 $U(q)$, i.e., consider the interacting part of the Hamiltonian in the form
\beq
{\cal H}_{int} = -\frac{1}{2} \sum U_{\alpha, \beta; \gamma,\delta} (k1,k_2;k_3,k_4) a^\dagger_{k_1,\alpha} a^{\dagger}_{k_2,\beta} a_{k_3,\gamma} a_{k_4,\delta}
\eeq
 where
 \beq
U_{\alpha, \beta; \gamma,\delta} (k1,k_2;k_3,k_4)= U(k_1-k_3) \delta_{\alpha \gamma} \delta_{\beta \delta}
\eeq
where $\alpha, \beta, \gamma, \delta$ are spin indices. We will not make any particular assumption about the form of $U(q)$, i.e., will not discuss the mechanism how it is obtained from the original Coulomb repulsion,
   except that we assume that $U(q)$ is analytic and the largest at $q=0$. In RG sense $U(q)$ can be understood as the effective four-fermion static interaction,
    obtained after integrating out high-energy fermions with energies between the bandwith $W$ and $\Lambda$, which is the fraction of the bandwidth, and which sets the upper energy cutoff for our low-energy theory.  The screening by high-energy fermions does not give rise to non-analyticities, hence it is safe to consider $U(q)$ as   an analytic function of $q$. We further assume that
$U(q)$ is small compared to $\Lambda$ and study the pairing within the perturbation theory (but not necessary the lowest order).

\begin{figure}
 \includegraphics[width = \columnwidth]{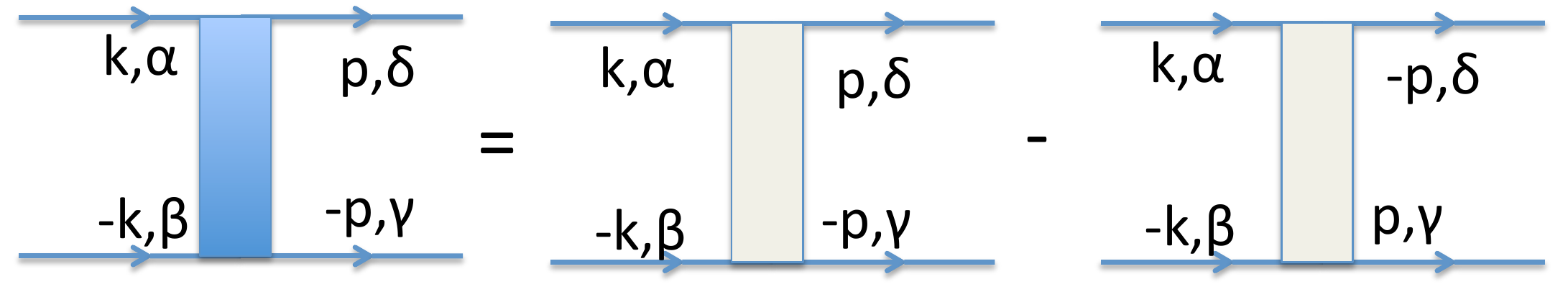}
 \caption{\label{fig:6}
 The vertex function, whose poles determine the pairing instability, is the fully renormalized and antisymmetrized interaction at zero total incoming and outgoing momenta.}
  \end{figure}

An instability in a particular pairing channel manifests itself through the appearance of a pole at zero frequency in the corresponding
 component of the vertex function $\Gamma_{\alpha, \beta; \gamma,\delta} (k_1,k_2;k_3,k_4)$ which describes two-particle collective
 bosonic excitations in a system of interacting fermions.  The vertex function incorporates multiple fermion-fermion scattering processes at energies below $\Lambda$, as well as the Pauli principle,  and constitutes the fully renormalized  and antisymmetrized four-fermion interaction (Fig.\ref{fig:6}.
 We first consider the pairing interaction to first order in $U$ and then add $U^2$ terms which will account for Kohn-Luttinger physics

 \subsection{First order in $U(q)$}
\begin{figure}
 \includegraphics[width=\columnwidth]{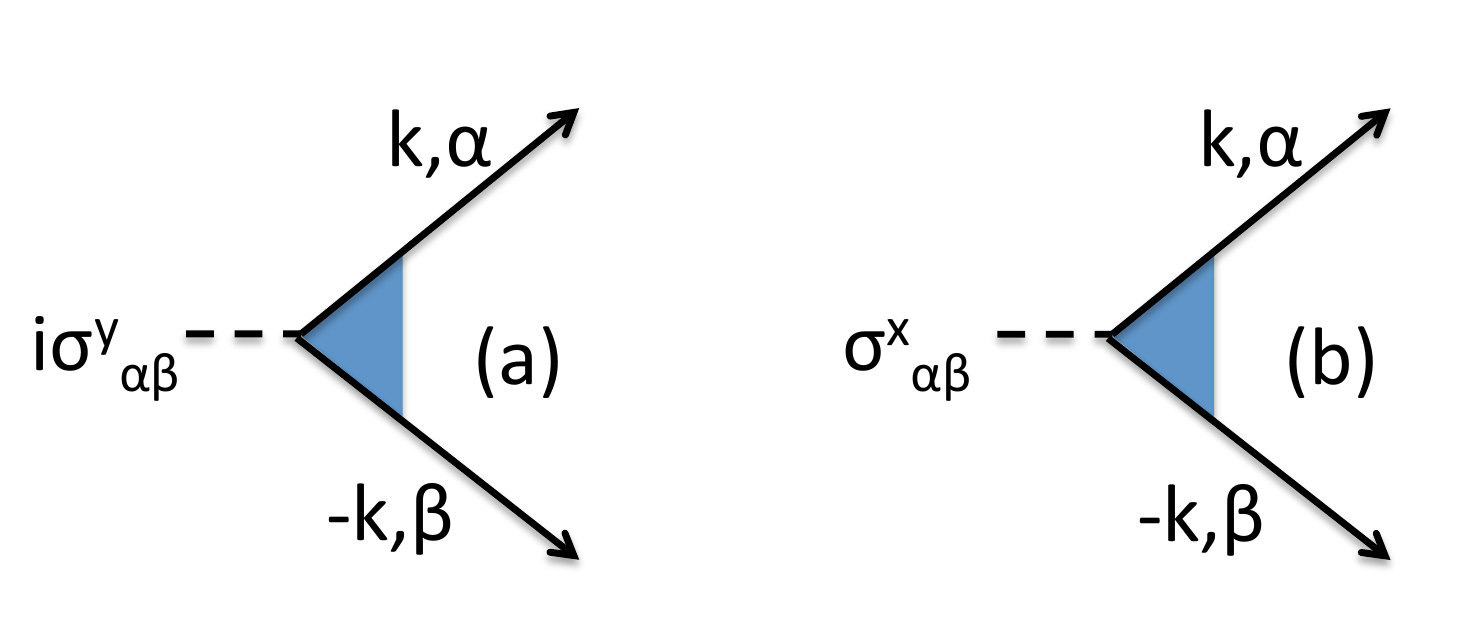}
 \caption{\label{fig:7}
 The pairing vertices for spin-singlet and spin-triplet channels.}
  \end{figure}
   To first order in $U(q)$, the vertex function coincides with the antisymmetrized  interaction
   \begin{widetext}
 \beq
\Gamma_{\alpha, \beta; \gamma,\delta} (k_1,k_2;k_3,k_4) = \frac{1}{2} \left(U(k_1-k_3) \delta_{\alpha \gamma} \delta_{\beta \delta} - U(k_1-k_4) \delta_{\alpha \delta} \delta_{\beta \gamma}\right)
\label{ch_2_1}
\eeq
At weak coupling,  the leading pairing instability occurs at a
 zero total momentum of a pair, and we set $k_1 = -k_2 = k$ and $k_3 = -k_4 = k'$. The part of $\Gamma$ responsible for the pairing is then
 \beq
\Gamma_{\alpha, \beta; \gamma,\delta} (k,k') = \frac{1}{2} \left(U(k-k') \delta_{\alpha \gamma} \delta_{\beta \delta} - U(k+k') \delta_{\alpha \delta} \delta_{\beta \gamma}\right)
\label{ch_2}
\eeq
The vertex function can be further decoupled into spin-singlet and spin-triplet channels as
\beq
\Gamma_{\alpha, \beta; \gamma,\delta} (k,k') = \frac{1}{4} \left(\left(U(k-k') - U(k+k')\right)
\left(\delta_{\alpha \gamma} \delta_{\beta \delta} + \delta_{\alpha \delta} \delta_{\beta \gamma}\right) +
\left(U(k-k') + U(k+k')\right)
\left(\delta_{\alpha \gamma} \delta_{\beta \delta} - \delta_{\alpha \delta} \delta_{\beta \gamma}\right) \right).
\label{ch_3}
\eeq

We now make use of the identities
\bea
&&\sum_{\gamma \delta} \left(\delta_{\alpha \gamma} \delta_{\beta \delta} - \delta_{\alpha \delta} \delta_{\beta \gamma}\right) \times \left(\delta_{\gamma \eta} \delta_{\delta \xi} - \delta_{\delta \eta} \delta_{\gamma \xi}\right)  = 2  \left(\delta_{\alpha \eta} \delta_{\beta \xi} - \delta_{\alpha \xi} \delta_{\beta \eta}\right) \nonumber \\
&& \sum_{\gamma \delta} \left(\delta_{\alpha \gamma} \delta_{\beta \delta} + \delta_{\alpha \delta} \delta_{\beta \gamma}\right) \times \left(\delta_{\gamma \eta} \delta_{\delta \xi} + \delta_{\delta \eta} \delta_{\gamma \xi}\right) = 2  \left(\delta_{\alpha \eta} \delta_{\beta \xi} + \delta_{\alpha \xi} \delta_{\beta \eta}\right) \nonumber \\
\label{ch_4}
\eea
\end{widetext}
to see that the singlet component of $\Gamma_{\alpha, \beta; \gamma,\delta} (k,k')$
 determines pairing instability in spin-singlet channel with the order parameter $\Delta_{\alpha,\beta} (k) = i\sigma^y_{\alpha\beta} \Delta (k)$,
  and the triplet component of $\Gamma_{\alpha, \beta; \gamma,\delta} (k,k')$
 determines pairing instability in spin-triplet channel with the order parameter $\Delta_{\alpha,\beta} (k) = i\sigma^y_{\alpha\gamma} \vec{\sigma}_{ \gamma \beta} \cdot \vec{d(k)} \Delta (k)$, where $\vec{d}(k)$ is a unit vector that determines the orientation of the triplet order parameter~\cite{volovik}.
 For definiteness, we set ${\vec d}(k)$ to be antiparallel to $z$ axis, then in spin-triplet channel $\Delta_{\alpha,\beta} (k) = \sigma^x_{\alpha\beta}
   \Delta (k)$. (see Fig. \ref{fig:7}).
    Accordingly,
      we introduce
\begin{eqnarray}
\Gamma^{s} = \frac{1}{2} \left(U(k-k') + U(k+k')\right),\nonumber \\
\Gamma^{t} = \frac{1}{2} \left(U(k-k') - U(k+k')\right),
\label{ch_5}
\end{eqnarray}

In rotationally-invariant systems, each of these two interactions can be further decoupled into orthogonal partial harmonics with either even angular momentum (for $\Gamma^s$) or odd angular momentum (for $\Gamma^t$). For lattice systems with non-circular FS, such simple decoupling is impossible and the maximum one can do is to decouple the interaction into a discrete set of representations for the corresponding space group.  The components from different sets are orthogonal to each other, but each set still contains an infinite numbers of eigenfunctions which are not orthogonal and do not decouple when we solve for superconducting $T_c$.

Such decoupling into different discrete sets turns out to be useful when all $k$ along the FS contribute about equally to the pairing.
 Although all eigenfunctions in each subset are technically of the same order, it turns out that the first harmonics
 contribute most, such that one can truncate the expansion by keeping only the lowest order terms, e.g., $\cos k_x - \cos k_y$ for d-wave pairing in the cuprates or
 $a \pm b \cos 2 \theta$ for the gaps on electron pockets in Fe-pnictides.  For fermions on a triangular lattice and $\mu$ near zero, the problem, however, is different because the the density of states along the FS is maximized   near van-Hove points, which then play the dominant role in the pairing (just like hot spots play the dominant role in the weak coupling description of the cuprates).  If one attempts to compare the contribution to the full $\Gamma$ from different eigenfunctions from the same subset, one finds that they all contribute nearly equally near such a `Van Hove' filling.

\begin{figure}
 \includegraphics[width = \columnwidth]{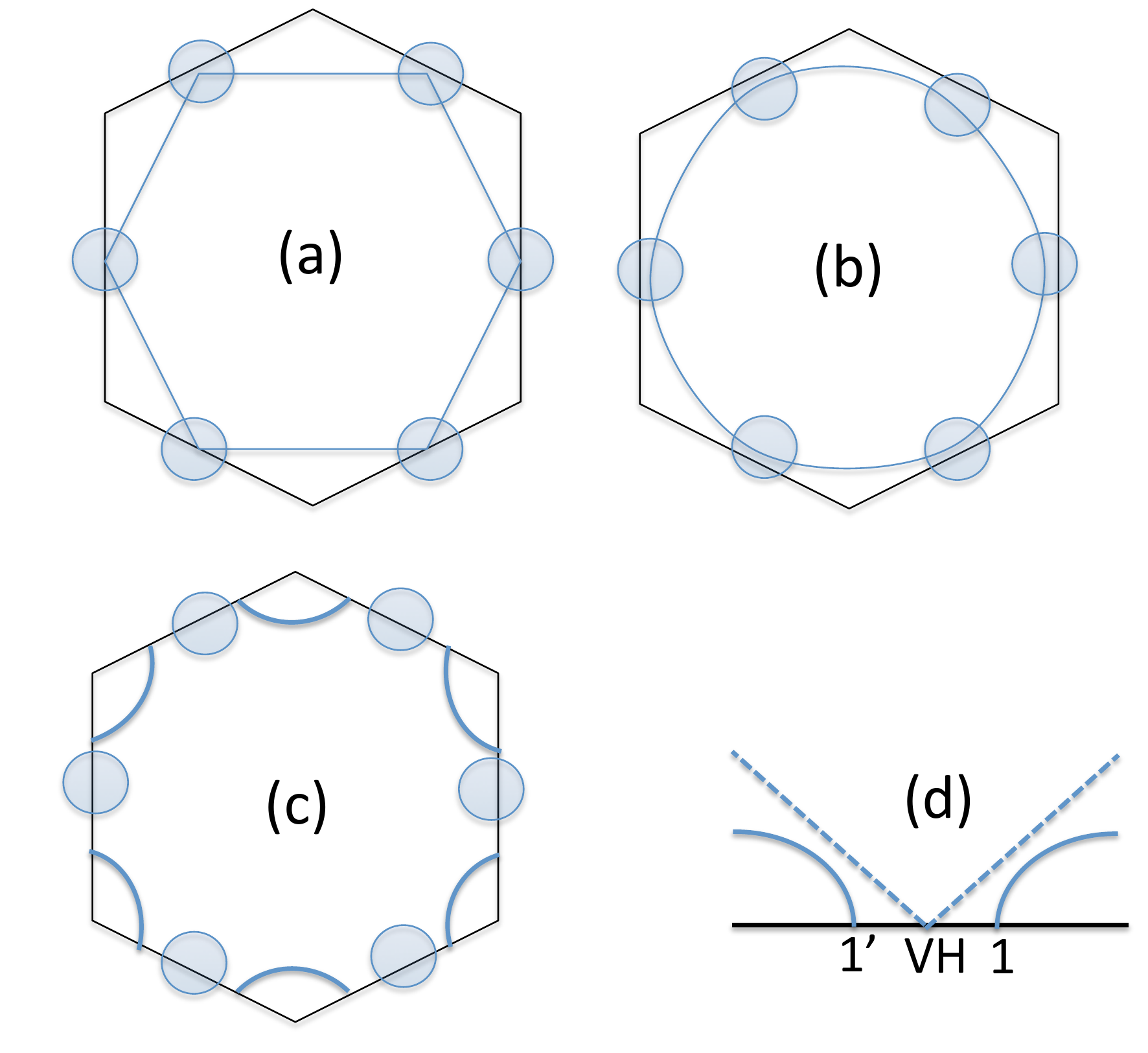}
 \caption{\label{fig:8}
 Patch model I: Assumes that the superconducting gap is concentrated in the shaded patches (a) Van-Hove doping. The patches are centered at van-Hove points. (b) and (c) -- above and below van-Hove doping, respectively.
 (d) -- for doping line in panel (c), the model allows for symmetric and anti-symmetric solutions between points $1$ and ${\bar 1}$.}
 \end{figure}

 \begin{figure}
 \includegraphics[width = \columnwidth]{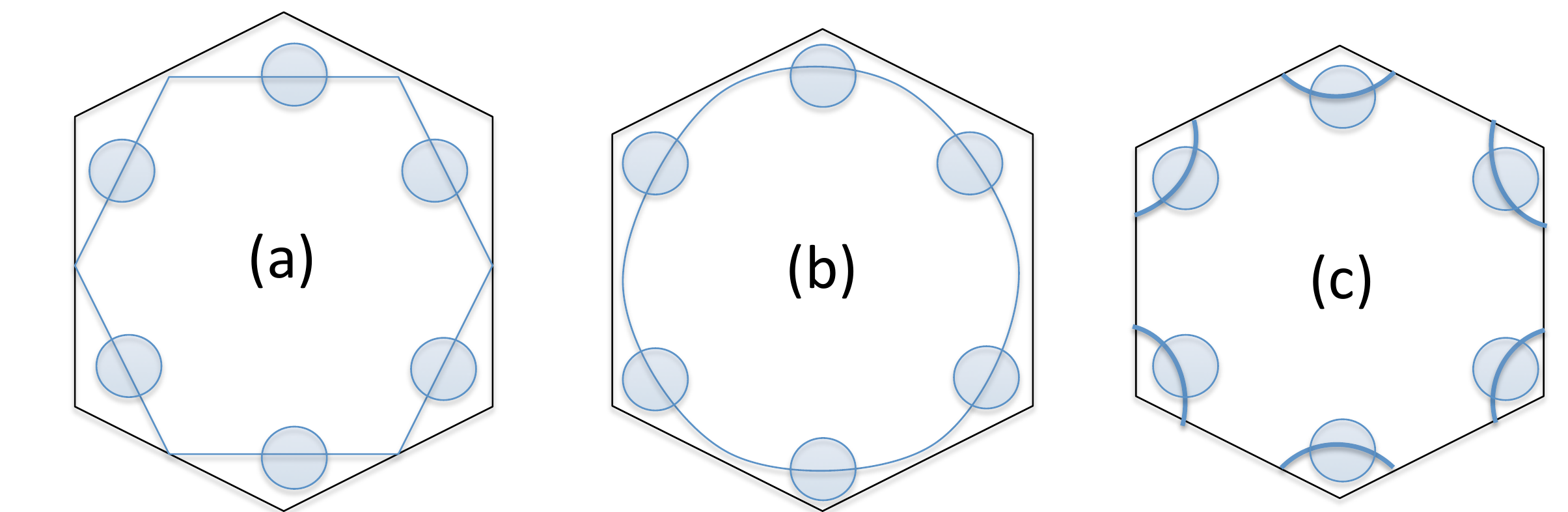}
 \caption{\label{fig:9}
 Patch model II. At Van-Hove doping (a) the patches are centered in between van-Hove points. (b) and (c) -- above and below van-Hove doping, respectively.
  For a FS in the form of disconnected pieces (panel c), the patches are located at the centers of FS arcs.}
 \end{figure}

  In this situation, it is advantageous to use a different approach and restrict the pairing problem to a discrete set of patches where the density of states is the largest and the dimensional coupling constant is the largest (see Figs. (\ref{fig:8}) and  (\ref{fig:9})).
  This is the approach which we take in this paper.  For fermions on a triangular lattice and near $\mu =0$, the density of states is maximized in the regions near van-Hove points.  There are six such regions (patches) on the FS (Fig. (\ref{fig:8})), but only three are independent because the other three are related to first three by inverting the sign of the momentum, and superconducting order parameters satisfy $\Delta (-k) = \Delta (k)$ for spin-singlet pairing and $\vec{d} (-k) = - \vec{d}(k)$ for spin-triplet pairing.  The three hot regions are all equivalent in the sense the $2{\bf k}_F$ for all three points differs by the reciprocal lattice vector and  the interaction at momentum ${\bf Q}$
   connecting any two different hot regions is the same $U(|{\bf Q}|)$.
      Accordingly, we define by $U(0)$  the direct interaction within any of the patches,  by $U(2k_F)$ the exchange (anti-symmetrized) part of intra-patch interaction with momentum transfer from ${\bf k}_F$ to $-{\bf k}_F$,  by $U(Q)$ the direct interaction between patches, and by $U(Q + 2k_F)$ the anti-symmetrized part of the inter-patch interaction. Because there are intra-patch and inter-patch interactions, we can introduce two different $\Gamma$ in either singlet or triplet channel:
      \bea
      &&\Gamma^{(1)}_s  = \Gamma_s (k,k) = \frac{1}{2} \left(U(0) + U(2k_F) \right)\nonumber \\
       &&\Gamma^{(2)}_s  = \Gamma_s (k,k+Q) = \frac{1}{2} \left(U(Q) + U(Q + 2k_F) \right)\nonumber \\
  &&\Gamma^{(1)}_t  = \Gamma_t (k,k) = \frac{1}{2} \left(U(0) - U(2k_F) \right)\nonumber \\
 &&\Gamma^{(2)}_t  = \Gamma_s (k,k) = \frac{1}{2} \left(U(Q) - U(Q + 2k_F) \right)
 \label{sa_1}
 \eea
We now introduce three order parameters $\Delta_i $, one for each patch, solve  $3\times 3$ pairing problem
 \bea
 &&\lambda_s \Delta_i = - \Gamma^{(1)}_s \Delta_i - \Gamma^{(2)}_s \sum_{k \neq i}\Delta_k \nonumber \\
  &&\lambda_t \Delta_i = - \Gamma^{(1)}_t \Delta_i - \Gamma^{(2)}_t \sum_{k \neq i}\Delta_k
  \label{sa_2}
\eea
   for singlet and triple channels, respectively, and obtain three eigenfunctions in each channel
   \bea
    \lambda^{(1)}_s &=& -\Gamma^{(1)}_s - 2 \Gamma^{(2)}_s, ~~  \lambda^{(2)}_s = \lambda^{(3)}_s = -\Gamma^{(1)}_s +  \Gamma^{(2)}_s \nonumber \\
    \lambda^{(1)}_t &=& -\Gamma^{(1)}_t - 2 \Gamma^{(2)}_t, ~~  \lambda^{(2)}_t = \lambda^{(3)}_t = -\Gamma^{(1)}_t +  \Gamma^{(2)}_t \nonumber \\
\label{sa_3}
\eea
For superconductivity, one needs at least one of these $\lambda$ to be positive.  If all $\lambda$ are either negative or zero, the normal state survives down to $T=0$.

 We will label this patch model as model I (Fig. (\ref{fig:8})).  We will also consider another patch model (patch model II), for which we we place patches in between van-Hove points (see Fig.(\ref{fig:9})).
 This second model  is less relevant at van Hove filling but it is useful at dopings when the FS splits into six disconnected segments because the patches in the model II are located at the centers of the FS segments.

\subsubsection{Extended Hubbard model}

We apply the results from the previous Section to the model with interaction $U(q)$, which extends up to second neighbors on a triangular lattice.
In momentum space
\begin{widetext}
\beq
U(q) = U_0 +U_1 \left( \cos{q_x} + 2 \cos{\frac{q_x}{2}} \cos{\frac{q_y \sqrt{3}}{2}} \right) + U_2
\left( \cos{q_y \sqrt{3}}  + 2 \cos{\frac{3 q_x}{2}} \cos{\frac{q_y \sqrt{3}}{2}} \right)
\label{sa_4}
\eeq
\end{widetext}
 where $U_0$, $U_1$, and $U_2$ are amplitudes of the on-site interaction and interaction between first and second neighbors, respectively.
  We assume that $U_0 \gg U_1 \gg U_2$, i.e., that the interaction $U(q)$ rapidly decreases at $q$ comparable to interatomic distances.

Using Eqs. (\ref{sa_1}) - (\ref{sa_4}), it is straightforward to obtain eigenfunctions in each of the  four channels at various fillings.
 Because $U(q)$ rapidly drops, we present only the leading contributions to various $\lambda_i$ (i.e., keep only the largest $U_i$). We will see, however, that at least in one case, $\lambda_i$ is only non-zero because of $U_2$.\\

1. {\it {Van-Hove filling, $\mu =0$.}}\\

For patch model I the three patches are centered at $k_1 = \left(0, -\frac{2\pi}{\sqrt{3}}\right)$, $k_2 = \left(\pi, \frac{\pi}{\sqrt{3}}\right)$, and
 $k_3 = \left(-\pi, \frac{\pi}{\sqrt{3}}\right)$.  The other three patches are centered at $-{\bf k}_i$.
 One can easily make sure that for each of these $k_i$,  $2 k_i \equiv 2k_F$ at van-Hove filling coincides with reciprocal lattice vector, the eigenfunctions in  spin-triplet channel vanish identically, i.e.,
 $\lambda^{(1)}_{s,I} = \lambda^{(2)}_{s,I} = \lambda^{(3)}_{s,I}$.
  In spin-singlet channel we have
 \beq
  \lambda^{(1)}_{s,I} = -3U_0 -U_1 - U_2,
  ~~  \lambda^{(2)}_{s,I} = \lambda^{(3)}_{s,I} = -4 \left(U_1 + U_2 \right)
  \label{sa_5}
  \eeq
  As long as all interactions are repulsive ($U_j >0$), both eigenvectors are negative, i.e., both spin-singlet channels  are repulsive.
  Note, however, that if on site repulsion is the strongest but first and second-neighbor interactions are attractive (the case of over-screening by high-energy fermions), the system becomes unstable towards $d-$wave superconductivity.

\begin{figure}
 (a) \includegraphics[width = 0.9 \columnwidth]{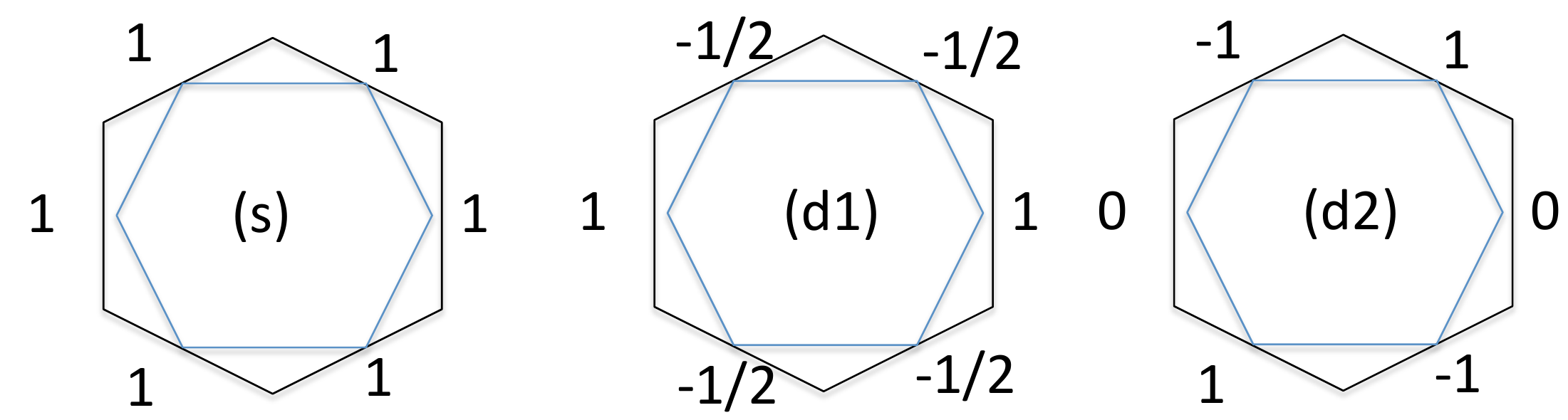}
(b)  \includegraphics[width = 0.9 \columnwidth]{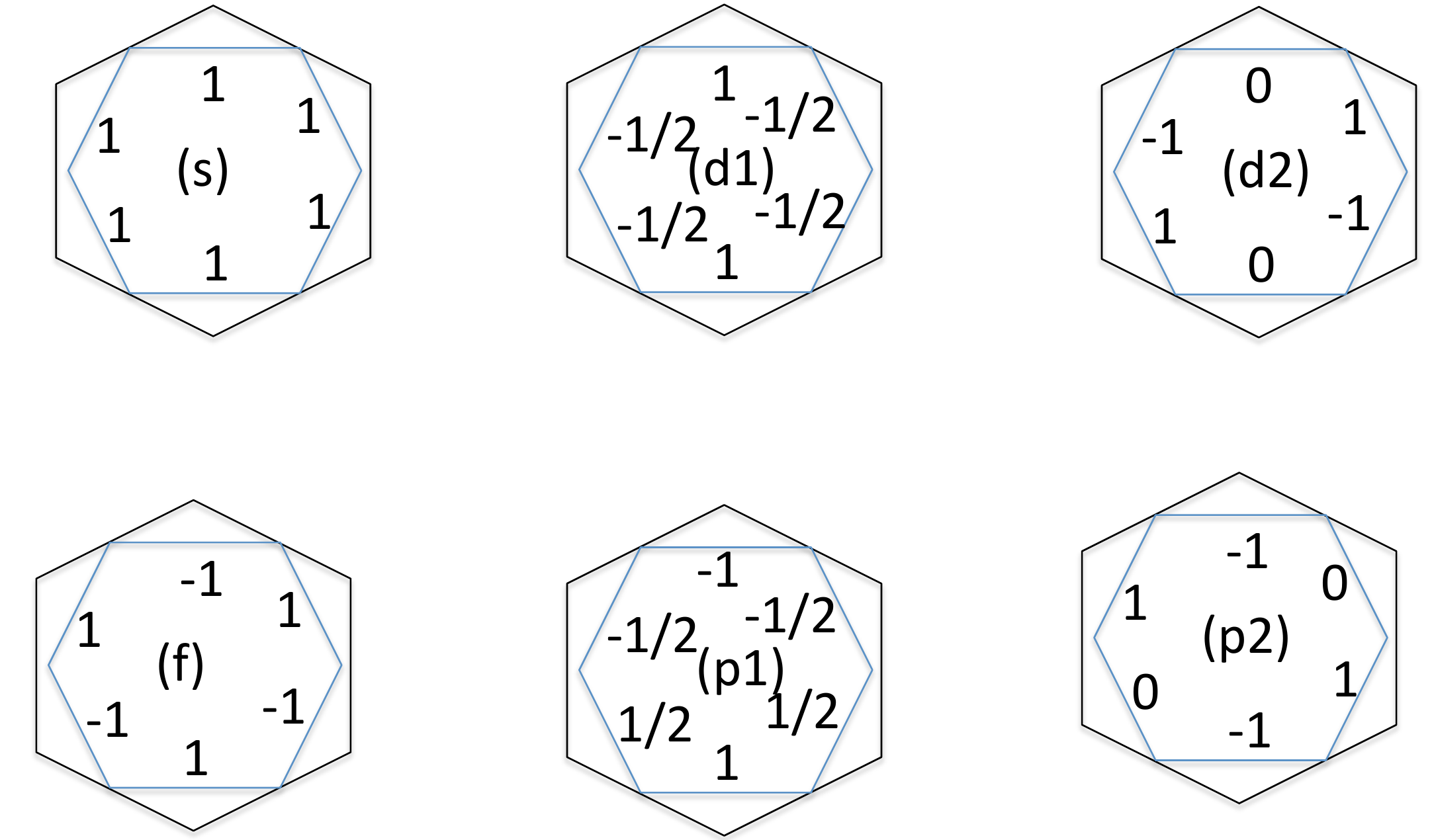}
\vspace{4mm}
(c) \includegraphics[width = 0.9 \columnwidth]{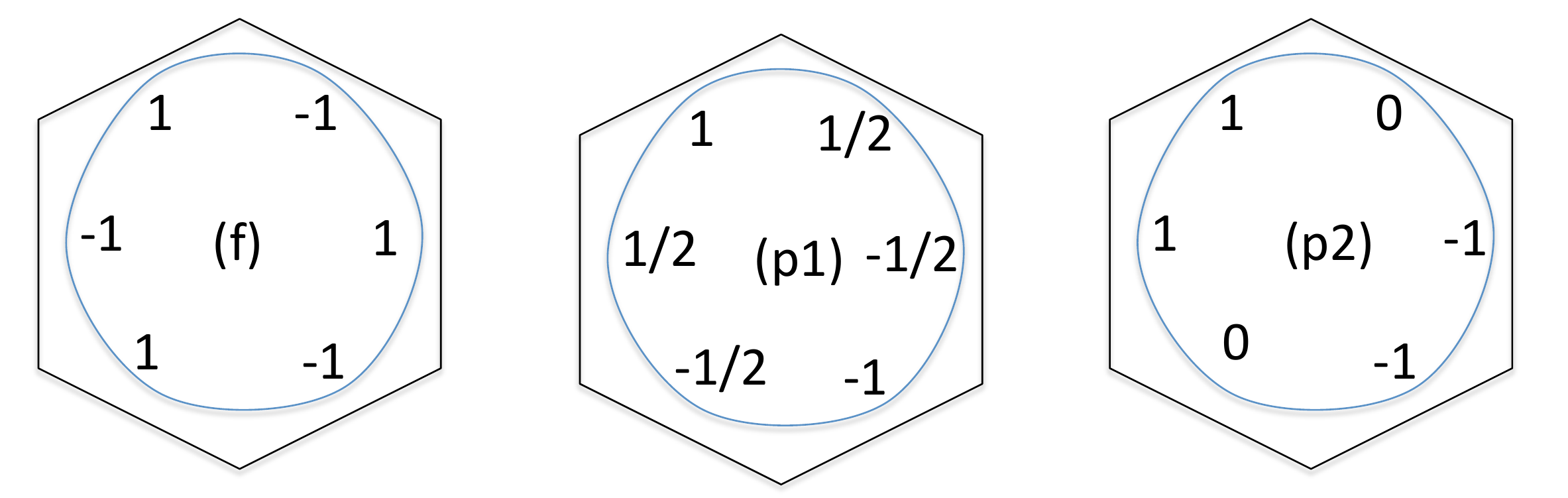}
 \caption{\label{fig:E1}
 (a) and (b) -- the structure of the superconducting gap in $s-$, $p-$, $d-$, and $f-$wave channels at van-Hove doping in patch models I and II respectively.
 In the $p-$wave and $d-$wave channels, the eigenvalues are doubly degenerate.  (c) The structure of the gaps in spin-triplet channels in patch model I at $\mu <0$,
  when the FS is singly-connected and does not reach van-Hove points.}
 \end{figure}

The structure of the superconducting gap is shown in Fig. \ref{fig:E1}a.
The eigenfunction $\lambda^{(1)}_s$ corresponds to an $s-$wave solution (all $\Delta_i$ are the same) , the eigenfunctions $\lambda^{(2)}_s$ and $\lambda^{(3)}_s$  describe two degenerate solutions which we term  $d-$wave ($d_1$ and $d_2$) because if we formally extend the gap to the full FS, we find that for the
 eigenfunctions, corresponding to $d_1$ and $d_2$, the gap changes sign two times as one circles the FS.
The gaps in the triplet channel vanish identically in patch model I together with eigenvalues $\lambda^{(i)}_{t,I}$, and we do not show them.

 For patch model II the three patches are centered at ${\bar k}_1 = - \left(\pi, 0\right)$, ${\bar k}_2 = \left(\frac{\pi}{2},
 \frac{\pi \sqrt{3}}{2}\right)$, and
 ${\bar k}_3 = \left(\frac{\pi}{2},
 -\frac{\pi \sqrt{3}}{2}\right)$.  The couplings are non-zero in both spin-singlet and spin-triplet channels. Setting $U_2=0$ for simplicity, we obtain
   \bea
  &&\lambda^{(1)}_{s,II} \approx -3 U_0,   \lambda^{(2)}_{s,II} =  \lambda^{(3)}_{s,II} = -U_1 \nonumber \\
   &&\lambda^{(1)}_{t,II} = - 4U_1,~~  \lambda^{(2)}_{t,II} =  \lambda^{(3)}_{t,II} =
   - U_1
    \label{sa_5_1}
   \eea
 We see that all bare couplings are repulsive for positive (repulsive)  $U_0$ and $U_1$.\\
We show the structure of the superconducting gaps (the eigenfunctions) in all six channels in Fig.  \ref{fig:E1}b.
In the spin-singlet sector,  eigenfunction  corresponding to $\lambda^{(1)}_{s,II}$ is obviously  an $s-$wave. The eigenfunctions corresponding to
$\lambda^{(2)}_{s,II} =  \lambda^{(3)}_{s,II}$ again  change sign two times as one circles the FS, and we keep calling them $d-$wave.
In the spin-triplet sector, the eigenfunction for $\lambda^{(1)}_{t,II}$ corresponds to an $f-$wave solution
 ($\Delta_i$ are the same for a triad separated by $Q$, but each $\Delta_i (-k) = -\Delta_i (k)$), such that the gap, if we extended it to the full
   FS, changes sign 6 times as one circles the FS.
  The eigenfunctions corresponding to  $\lambda^{(2)}_{t,II}$ and $\lambda^{(3)}_{t,II}$  change sign two times,
   as one circle the FS, and we label them as $p-$wave solutions.\\

{\it {2.  A smaller filling, $\mu <0$}}\\

We consider relatively small deviations from van-Hove filling, at which patch models still make sense (see Fig.(\ref{fig:8})).
For model I, the three patches are now centered at
$k_1 = \left(0, -\frac{2}{\sqrt{3}} (\pi - \delta)\right)$, $k_2 = \left(\pi -\delta, \frac{\pi -\delta}{\sqrt{3}}\right)$, $k_3 = \left(-(\pi -\delta), \frac{\pi - \delta}{\sqrt{3}}\right)$, where $\delta \ll 1$ and $\mu \approx -2 t \delta^2$. Once $\delta$ is non-zero, the couplings in spin-triplet channel become finite. We have
 \beq
  \lambda^{(1)}_{t,I} = -16 U_2 \delta^2, ~~  \lambda^{(2)}_{s,I} = \lambda^{(3)}_{s,I} = -\left(3 U_1 + U_2 \right)\delta^2
  \label{sa_6}
  \eeq
As long as $U_1$ and $U_2$ are positive (repulsive), the interactions in both spin-singlet channels are repulsive, i.e, spin-triplet superconductivity does not emerge.
 The interactions in spin-singlet channels were repulsive for $\delta =0$ (for $U_i >0$) and remain so at a non-zero positive $\delta$.
The structure of the superconducting gaps in spin-singlet channel does not change from Fig. \ref{fig:E1}a, but now there appear  eigenfunctions in $f-$wave and $p-$wave, corresponding to the
eigenvalues in  Eq. (\ref{sa_6}).  We show these eigenfunctions in Fig. \ref{fig:E1}c

  For model II (Fig.(\ref{fig:9})) the patches are centered at ${\bar k}_1 = -(\pi - {\bar \delta}, 0), ~ {\bar k}_2 = (\frac{\pi + {\bar \delta}}{2}, \frac{\sqrt{3}}{2} (\pi + {\bar \delta})),~ {\bar k}_3 = (\frac{\pi + {\bar \delta}}{2}, -\frac{\sqrt{3}}{2} (\pi + {\bar \delta}))$ and at $-{\bar k}_i$, where ${\bar \delta} = -2 \arcsin{[(\sqrt{(t-\mu)/t}-1)/2]}$.  For $\mu <0$, ${\bar \delta}$ is negative.
   The couplings are given by (setting $U_2 =0$)
   \begin{widetext}
 \bea
  &&\lambda^{(1)}_{s,II} \approx -3 U_0,   \lambda^{(2)}_{s,II} =  \lambda^{(3)}_{s,II} = -\frac{U_1}{2} \left(2 + \cos{2 {\bar \delta}} - \cos{\bar \delta} \left(1 +
   4 \sin{{\bar \delta}/2}\right)\right)    \nonumber \\
   &&\lambda^{(1)}_{t,II} = - 4U_1 \cos^2{{\bar \delta}/2} \left(1 + \sin{{\bar \delta}/2}\right)^2,~~  \lambda^{(2)}_{t,II} =  \lambda^{(3)}_{t,II} =
   - U_1  \cos^2{{\bar \delta}/2} \left(1 -2 \sin{{\bar \delta}/2}\right)^2
    \label{sa_9}
   \eea
   \end{widetext}
  For ${\bar \delta} =0$ we reproduce (\ref{sa_5_1}).  We see that all interactions remain repulsive for arbitrary ${\bar \delta}$, as long as $U_0$ and $U_1$ are positive.  The structure of the eigenfunctions in $s, d$, and $p-$channels remains the same as in Fig. \ref{fig:E1}b.\\

 {\it {3. A larger filling, $\mu >0$.}}\\

\begin{figure}
 \includegraphics[width = \columnwidth]{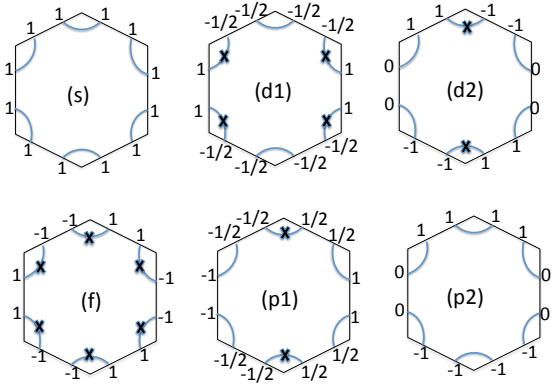}
 \caption{\label{fig:10}
 The structure of superconducting gaps in $s-$wave, $f-$wave, and two degenerate $d-$wave and $p-$wave channels for the symmetric solution for the patch model I
  for $\mu >0$, when the FS consists of six disconnected pieces. About the same solutions are obtained in patch model II in $s-$wave and $p-$wave channels. }
 \end{figure}

\begin{figure}
 \includegraphics[width = \columnwidth]{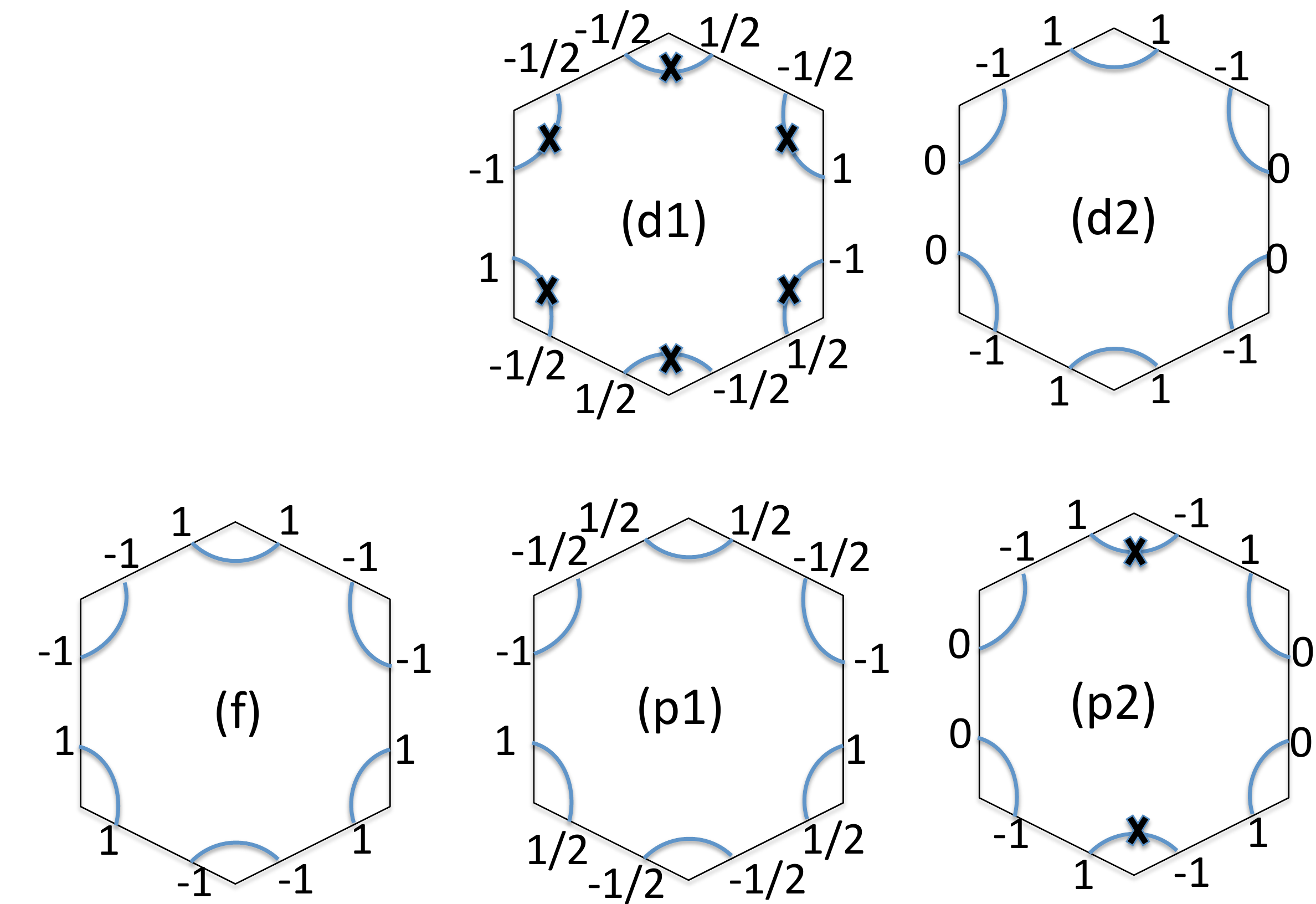}
 \caption{\label{fig:11}
 The same as in Fig. \protect\ref{fig:10}, but for the antisymmetric solution for the patch model I.
 About the same solutions are obtained in patch model II in $f-$wave and $d-$wave channels.
 }
 \end{figure}

 The situation becomes more interesting at a larger filling, when the FS splits into 6 pockets, each centered at the corner of the  BZ.  At small positive $\mu = 2 t \delta^2$ one can still focus on patch model I with patches near Van Hove points  $\left(0, -\frac{2\pi}{\sqrt{3}}\right)$, $\left(\pi, \frac{\pi}{\sqrt{3}}\right)$, and $\left(-\pi, \frac{\pi}{\sqrt{3}}\right)$, but now each van-Hove point splits into two FS points, as shown in Fig... For example,
  $k_1 = \left(0, -\frac{2\pi}{\sqrt{3}}\right)$ splits into  $k_{1,+} = \left(2\delta, -\frac{2\pi}{\sqrt{3}}\right)$ and  $k_{1,-} = \left(-2\delta, -\frac{2\pi}{\sqrt{3}}\right)$.  This splitting opens a possibility to consider two types of solutions for the gaps in each patch: a symmetric one, for which $\Delta_{k_{1,+}} =  \Delta_{k_{1,-}}$ and antisymmetric one, for which $\Delta_{k_{1,+}} = - \Delta_{k_{1,-}}$. We label the corresponding $\lambda$ with additional sub-indices $s$ and $a$.

  A simple analysis shows that the expressions for the vertices in  spin-signet and spin-triplet channels for symmetric and antisymmetric solutions  are obtained by replacing in Eq. (\ref{sa_1})
  \bea
  &&U(0) \to \frac{U(0) \pm U(k_{1,+} - k_{1,-})}{2},\\
 &&  U(2k_F) \to \frac{U(2k_{1+}) \pm U(k_{1,+} + k_{1,-})}{2}, \nonumber\\
  &&U(Q) \to \frac{U(k_{1,+} - k_{2,+}) \pm U(k_{1,+} - k_{2,-})}{2},\nonumber\\ && U(Q +2k_F) \to \frac{U(k_{1,+} + k_{2,+}) \pm U(k_{1,+} + k_{2,-})}{2},\nonumber
  \label{sa_9_1}
  \eea
  where upper (lower) sign is for symmetric (antisymmetric) solution.

For the symmetric solution we have, formally keeping $\delta$ as arbitrary number between $\delta =0$ and $\delta =\pi/3$, at which  the pockets disappear, and neglecting $U_2$,
  \bea
  &&\lambda^{(1,s)}_{s,I} \approx -3 U_0,   \lambda^{(2,s)}_{s,I} =  \lambda^{(3,s)}_{s,I} = - U_1 (\cos{\delta} + \cos{2\delta})^2 \nonumber \\
   &&\lambda^{(1,s)}_{t,I} = 0,\nonumber\\
   &&  \lambda^{(2,s)}_{t,I} =  \lambda^{(3,s)}_{t,I} = 0
   \label{sa_7}
   \eea
   where $\mu = 4t \cos{\delta} (1 - \cos{\delta})$.
 Symmetric
  solutions for $\Delta_i$ in various channels for the model I are shown in Fig.\ref{fig:10}.  For $f-$wave solution, there are six zeros, one on each disconnected segment of the FS. For each $d-$wave solution there are four nodes. For $p-$wave solutions, there are two nodes.
   The signs of $\lambda$'s in (\ref{sa_7}) are, however, such that at this level  no superconductivity emerges down to $T=0$ when $U_0$ and $U_1$ are positive.

 For the antisymmetric solution we obtain, again neglecting $U_2$
   \bea
  &&\lambda^{(1,a)}_{s,I} = 0,   \lambda^{(2,a)}_{s,I} =  \lambda^{(3,a)}_{s,I} = 0 \nonumber \\
   &&\lambda^{(1,a)}_{t,I} = - 4U_1 \sin^2{\delta} \left(1 + \cos{\delta}\right)^2,\nonumber\\
   &&   \lambda^{(2,a)}_{t,I} =  \lambda^{(3,a)}_{t,I} = - U_1 \sin^2{\delta} \left(1 -2 \cos{\delta}\right)^2
    \label{sa_8}
   \eea
 The structure of $\Delta_i$ in various channels is shown in Fig.\ref{fig:11}. We keep calling the solutions $s, p, d$ and $f-$wave, although the number of zeros on the FS now depends on whether the solution is symmetric or antisymmetric.  In particular, antisymmetric $f-$wave solution has no nodes on the FS.
 (the nodes fall in between the Fermi pockets), while one of antisymmetric $d-$wave solutions has nodes on each of six disconnected segments of the FS.

   Like for the symmetric case, the couplings in (\ref{sa_8}) are  either repulsive or zero.
In the formal limit $\delta = \pi/3$, when FS segments shrink into points, we have negative $\lambda^{(1,a)}_{t,I} = - 27 U_1/4$ for $f-$wave channel,   however the couplings in $s-$wave, $p-$wave and $d-$wave channels all vanish.

 For  patch model II, we have
  \begin{widetext}
 \bea
  &&\lambda^{(1)}_{s,II} \approx -3 U_0,   \lambda^{(2)}_{s,II} =  \lambda^{(3)}_{s,II} = -\frac{U_1}{2} \left(2 + \cos{2 {\bar \delta}} - \cos{\bar \delta} \left(1 +
   4 \sin{{\bar \delta}/2}\right)\right)    \nonumber \\
   &&\lambda^{(1)}_{t,II} = - 4U_1 \cos^2{{\bar \delta}/2} \left(1 + \sin{{\bar \delta}/2}\right)^2,~~  \lambda^{(2)}_{t,II} =  \lambda^{(3)}_{t,II} =
   - U_1  \cos^2{{\bar \delta}/2} \left(1 -2 \sin{{\bar \delta}/2}\right)^2
    \label{sa_9_1_1}
   \eea
   \end{widetext}
 as in Eq. (\ref{sa_9}), but now $\mu >0$ and ${\bar \delta} = 2 \arcsin{[(1 - \sqrt{(t-\mu)/t})/2]}$ is positive.
  At ${\bar \delta} = \pi/3$, $\mu =t$, and FSs disappear.

\begin{figure}
 \includegraphics[width = \columnwidth]{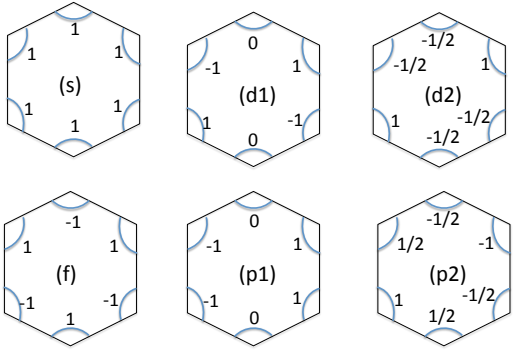}
 \caption{\label{fig:E2}
 The structure of the gap in $s$, $p$, $d$, and $f-$wave channels in patch model II at $0 <\mu <t$, when the FS consists of six disconnected segments.}
 \end{figure}
 We show the structure of eigenfunctions in various channels in Fig.\ref{fig:E2}.
 Comparing these solutions with those in patch model I and assuming that the system allows only symmetric or antisymmetric solutions with respect to former position of van Hove points,  we realize that the $d-$wave and $f-$wave eigenfunctions in patch model II (corresponding to $\lambda^{(1)}_{t,II}$ and $\lambda^{(2,3)}_{s,II}$)
  have the same structure as antisymmetric solutions in patch model I (the ones for $\lambda^{(1,a)}_{t,I}$ and $\lambda^{(2,a)}_{s,I} =\lambda^{(2,a)}_{s,I}$), while
 the eigenfunctions in  $s-$wave and $p-$wave channels in patch model II (corresponding to $\lambda^{(1)}_{s,II}$ and $\lambda^{(2,3)}_{t,II}$)
  have the same structure as symmetric solutions in patch model I (the ones for $\lambda^{(1,s)}_{s,I}$ and $\lambda^{(2,s)}_{t,I} =\lambda^{(2,s)}_{t,I}$).
 The eigenvalues are again generally negative (repulsive), if $U_0$ and $U_1$ are positive, but vanish in $d-$wave and $p-$wave channels when ${\bar \delta} = \pi/3$, i.e., when the size of a FS shrinks to zero.

 The conclusion of the analysis in this subsection is that there are multiple choices for the the structure of superconducting gap for fermions on a triangular lattice. The "basic" symmetry of the gap function is either $s$, or $p$, or $d$, or $f-$wave, but the structure of the gap in each representation depends on the choice of
  where the gap is maximized, what in our study implies the choice of the patch model.  The eigenvalues in various channels depend on the coordinate dependence of the screened Coulomb interaction $U(r)$. For repulsive interaction $U(r)$, which extends to second nearest neighbors, we found that the bare couplings in all pairing channels (i.e.,
  the couplings to order $U$) are repulsive, but some are either zero or close to zero.
 The coupling in a conventional spin-singlet $s-$wave channel is repulsive and of order of the on-site interaction $U_0$. The couplings in other channels
  vanish in the Hubbard model with on-site interaction but are generally non-zero and repulsive in an extended Hubbard model which includes interaction between nearest and second nearest neighbors.  In particular, in patch model I, the couplings in spin-singlet channel are repulsive at the van Hove filing ($\mu =0$), but the ones in spin-triplet channel are strictly zero.  At smaller fillings (negative $\mu$), bare couplings in all channels are repulsive.
  At larger fillings, when the FS is separated into six disconnected segments, the bare $f-$wave coupling is zero for the the symmetric solution, and the bare d-wave coupling is zero for
    the antisymmetric solution. For this last solution, the coupling in $p-$wave channel gets smaller near full filling. In patch model II the bare couplings are also
     repulsive a a generic filling, but the ones in $d-$wave and $p-$wave channels get smaller when the size of the FS shrinks.

\subsection{Kohn-Luttinger renormalizations}

 We now extend this analysis to second order in $U$ and include the renormalization of the pairing interaction by fermions with energies below the theory cutoff $\Lambda$. As noted in the introduction, such analysis has been first performed by Kohn and Luttinger (KL) for a 3D Fermi liquid with a generic short-range interaction, and second-order contributions to the pairing interactions are often called KL contributions.  We will use this notation.

   One can argue that  KL renormalization is relevant at weak coupling only in a situation when the bare interaction is zero, otherwise
    there is a repulsion to first order in $U(q)$ and KL terms cannot overcome it.  This is generally true if $U(q)$ is asymptotically small.
  However, we will see that attractive contributions to $\lambda^{(2,s)}_{s,I} = \lambda^{(3,s)}_{s,I}$ and $\lambda^{(1,a)}_{t,I}$
   (symmetric $d-$wave and  antisymmetric $f-$wave, respectively) appear at the second order in the on-site Hubbard interaction $U_0$. The first-order terms in these channels are repulsive, but are of order $U_1$  If, as it often happens, screened Couplomb interaction $U(r)$ rapidly drops with increasing $r$, we have $U_0 >> U_1$. In this situation,
 the second order contribution in $U_0$ becomes comparable to  $O(U_1)$ already within weak coupling, when perturbative approach is still valid (e,g., $U^3_0$ terms are small compared to KL terms).

In the rest of this section we compute $O(U^2)$ terms near but still at some distance from van Hove filling. Right at van-Hove filling, the KL
  $U^2$ terms  contain logarithms which can only be cut by external temperature, and higher-order terms in $U$ contain higher powers of divergent logarithms.
   In this situation,  one cannot restrict the analysis to terms second order in $U$ and has to sum up an infinite series of terms.   We discuss van-Hove filling in more detail later in the paper, in Sec. 4. Here and in the next Section we assume that logarithms are cut by a non-zero $|\mu|$ and $U(q)$ is small enough such that perturbation theory is valid

 The computation of KL renormalization of the pairing interaction is straightforward and has been discussed several times in the literature. One has to dress the bare interaction by the renormalization in the particle-hole channel. Approximating $U(q)$ by $U_0$, one finds that only exchange diagram in Fig.\ref{fig:1} contributes. In this situation, the renormalized interactions are
 \begin{widetext}
 \beq
U_{ren}(k-k') = U(k-k') + U^2_0 \Pi (k+k'),  U_{ren}(k+k') = U(k+k') + U^2_0 \Pi (k-k')
\label{sa_10}
\eeq
\end{widetext}
 where $\Pi (q) = \int (d^2 l d \omega/(8 \pi^3) G(k, \omega) G(l+q, \omega)$ is the particle-hole polarization bubble (defined with a positive sign). For a circular FS in 2D, $\Pi (q <2k_F) = m/(2\pi)$.

To understand what KL terms do, it is sufficient to consider the two limiting cases: (i) the case $\mu \approx t$, when disconnected FS segments are about to disappear,
 and (ii) a small deviation from van Hove point. To keep presentation focused, we will not consider in detail how KL renormalization affects all possible channels, and instead focus on the
  two most obvious choices -- a spin-triplet nodeless $f-$wave state near $\mu \approx t$ and doubly degenerate $d-$wave state near $\mu =0$. We show that for both states, KL terms are attractive, of order $U^2_0$, and well may overshoot repulsive terms of order $U_1$.
 We checked KL renormalization in other channels and found that they are not competitors to the two which we discuss below.

  \subsubsection{Near full filling.}

 Near full filling, the most straightforward approach is to consider patch model II in which patches are centered in the middle of each FS segment.

At $\mu  \leq t$, one can easily make sure that there are only two relevant interactions $U_{ren} (0)$ and $U_{ren} (2k_F)$, as if we define the corner BZ points as $k_1 =
  (-\frac{4\pi}{3},0)$, $k_2 = (\frac{2\pi}{3}, \frac{3\pi}{\sqrt{3}})$ and $k_3 = (\frac{2\pi}{3}, \frac{3\pi}{\sqrt{3}})$, then $k_i-k_j$ is zero up to reciprocal lattice vector and $k_i + k_j = 2k_i$, again up to a reciprocal lattice vector.  Then all we need to is to compute $\Pi(0)$ and $\Pi (2k_F)$.

  The fermionic dispersion  within each segment can be obtained by expanding near near the top of the band, i.e., for $k = (-\frac{4\pi}{3} + l_x, l_y)$,
  the dispersion
  \beq
  \epsilon (k) = (t-\mu) - \frac{3 t}{4} l^2 - \frac{\sqrt{3} t}{8} l_x \left( l^2_x - 3 l^2_y\right) + ...
  \label{sa_11}
  \eeq
  Evaluating the particle-hole bubbles $\Pi (0)$ and $\Pi (2k_F)$, we obtain that for both the momentum integrals are confined to small ${\bf l}$ when Eq. (\ref{sa_11}) is valid. The results are
  \bea
  &&\Pi (0) \approx \frac{1}{\sqrt{3}} \frac{1}{\pi t},\nonumber \\
   &&\Pi (2k_F) = \frac{1}{2}\Pi (0)  + \frac{\sqrt{3}}{54 \pi t} \left(\frac{t-\mu}{t}\right) \approx  \frac{1}{2}\Pi (0)
  \label{sa_12}
  \eea
  The approximate factor of 2 difference between $\Pi (0)$ and $\Pi (2k_F)$ can be easily understood.  Indeed, $\Pi (0)$ is the sum of contributions from all six segments of the FS. Its independence on $t-\mu$ is the known result in 2D: after the frequency integration in $
  \Pi (0) = \lim_{q \to 0} \int d^2 l d \omega G(l, \omega) G(l+q, \omega)$ the smallness of the phase space precisely cancels out by the smallness of the energy $\epsilon_l$ in the denominator of the
       For $\Pi (2k_F)$ $\epsilon (l)$ and $\epsilon (l+2k_F)$ are simultaneously small only in three FS segments out of six. In the other three,  if we choose $l$ to be on the FS, $\epsilon (l_F+2k_F)$ will be far from zero. For example, if we choose $2k_F = (-\frac{8\pi}{3},0)$, then $\epsilon (l)$ and $\epsilon (l+2k_F)$ are both small for $l$ near $(\frac{4\pi}{3},0)$, $(-\frac{2\pi}{3},\frac{2\pi}{\sqrt{3}})$, and $(-\frac{2\pi}{3},-\frac{2\pi}{\sqrt{3}})$ while for $l$  near
      $(-\frac{4\pi}{3},0)$, $(\frac{2\pi}{3},\frac{2\pi}{\sqrt{3}})$, and $(\frac{2\pi}{3},-\frac{2\pi}{\sqrt{3}})$, $\epsilon_l$ is small, but $\epsilon (k+2k_F) \approx -9t$.  The contribution from these three corners is then small in $t-\mu$ (the second term in $\Pi (2k_F)$ in (\ref{sa_12})).

   Using $\Pi (2k_F) \approx (1/2) \Pi (0)$, we obtain, keeping only $U_0$ term
    \bea
&& U_{ren}(0) = U_0 + \frac{1}{2} U^2_0 \Pi (0), \nonumber \\
&&  U_{ren}(2k_F) = U_0 + U^2_0 \Pi (0),  \nonumber \\
~~\Pi_0 = \frac{1}{\sqrt{3} \pi t}
\label{sa_14}
\eea
 Substituting these $U_{ren}$ into spin-singlet and spin-triplet components of the vertex function and re-evaluating $\lambda_i$ in different channels we
  find after simple algebra that the KL term gives a  positive (attractive) contribution
   of order $U^2_0 \Pi(0)$ to $f-$wave coupling, which becomes
     \beq
\lambda_f = = \lambda^{(1)}_{t,II}  + \frac{3}{4} U^2_0 \Pi(0) \approx - \frac{27}{4} U_1 + \frac{\sqrt{3}}{4\pi} \frac{U^2_0}{t}
    \label{sa_15}
   \eeq
  We see that the interaction in the  $f-$wave channel in the patch model II (same as in the antisymmetric $f-$wave channel in patch modrel I) becomes attractive if second-order KL contribution from the on-site interaction $U_0$ exceeds the repulsive first-order contribution from nearest-neighbor interaction $U_1$. We remind that the antisymmetric $f-$wave solution is the one in which the gap does not change sign along a given FS segment but changes sign between nearest FS segments.

Suppose that  the KL contribution is larger and the interaction in f-wave channel is attractive.  We then have
$T_{c,f} \sim \epsilon_F e^{-1/{\bar \lambda}_f}$, where $\epsilon_F \sim t-\mu$, and ${\bar \lambda} = \lambda_f/(3\sqrt{3} \pi t)$ is the dimensionless coupling.
We have,  without $U_1$ term,
${\bar \lambda}_f = (1/12 \pi^2) (U_0/t)^2 \approx 0.68 (U_0/W)^2$, where $W =9t$ is the bandwidth.
  Note that $T_c$ vanishes at $\mu = t$ due to vanishing prefactor.  However the coupling ${\bar \lambda}_f$ remains finite in this limit.
 This last result is a peculiarity of 2D where the density of states on the FS does not depend on the value of Fermi momentum.

One can easily extend the analysis to order $O(t-\mu)$ and analyze how $T_{c,f}$ evolves with decreasing $\mu$. We found that the $f-$wave coupling {\it increases}
 with increasing fermionic density. Explicitly,
 \beq
 {\bar \lambda}_f = \frac{1}{12\pi^2} \left(\frac{U_0}{t}\right)^2 \left[ 1 + \frac{5}{9} \frac{t-\mu}{t}\right]
 \label{sa_17}
 \eeq
 As a result, $f-$wave pairing gets stronger as one moves away from full to van Hove filling (see Fig. \ref{fig:13}a

For completeness and for comparison of the couplings between $f-$wave and $d-$wave channels later in the paper we
  computed KL renormalization of the coupling $\lambda^{(2,s)}_{s,I}$ for the symmetric $d-$wave channel in patch model I.
  We found that for this coupling, which we label as $\lambda_d$ to make presentation more simple,   KL terms make initially repulsive interaction even more repulsive.
 Specifically,
\beq
\lambda_d = \lambda^{(2,s)}_{s,I} -\frac{3}{4} U^2_0 \Pi (0) \approx - \frac{\sqrt{3}}{4\pi} \frac{U^2_0}{t} <0
    \label{sa_16}
   \eeq

\subsubsection{Near Van Hove filling}

\begin{figure}
 \includegraphics[width = \columnwidth]{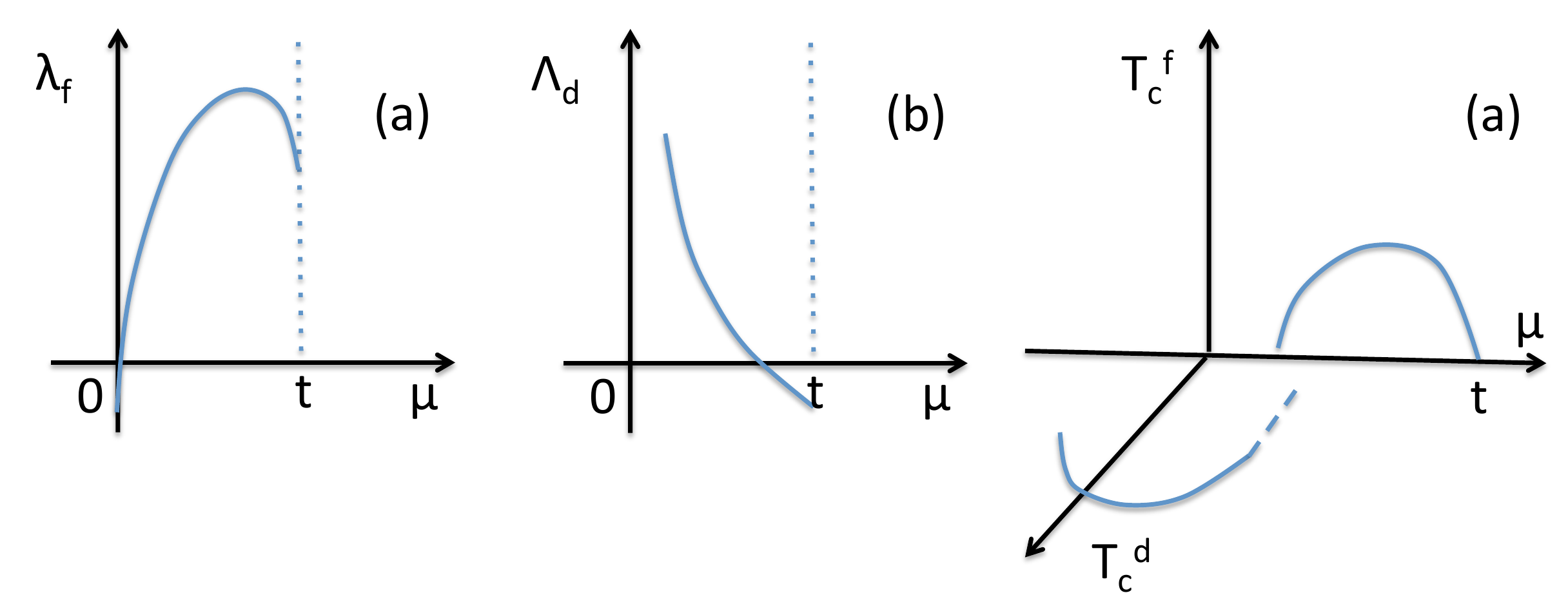}
 \caption{\label{fig:13}
 (a) and (b) The coupling constants in $f-$wave channel and in $d-$wave channel as functions of the chemical potential for fermions on a triangular lattice.  (c). The transition temperatures for $f-$wave and chiral $d+id$ superconductivity.  The coupling in $f-$wave channel remains finite and positive at   $\mu =t$, when the FS disappears, but
  $T_c$ for $f-$wave vanishes because of vanishing prefactor. Near van-Hove doping, $f-$wave channel is already repulsive. As a result, $T_c$ for $f-$wave is peaked in between $\mu =0$ and $\mu =t$.   The degenerate $d-$ wave channel has the largest coupling at van-Hove doping, but becomes repulsive at $\mu \approx t$. Whether the attractive regions in $d-$wave and $f-$wave channels overlap depends on the interplay between KL terms and the first-order terms coming from interaction between first, second, and, possibly, further neighbors. Moreover, at intermediate $\mu$, SDW well may become the leading instability.
  }
 \end{figure}

Continue first with the coupling in the $f-$channel in patch model II.
Near van Hove doping, the expression for $\lambda_f$ is
\begin{widetext}
\beq
\lambda_f = \lambda^{(1)}_{t,II} + \frac{U^2_0}{2} \left[\Pi (0) + 2 \Pi (Q) - \Pi (2k_F) - 2 \Pi (Q + 2k_F)\right]
\label{sa_18}
\eeq
\end{widetext}
For $\mu \approx 0$  all four $\Pi$'s are different and have contributions from low-energy fermions from different numbers of patches.
$\Pi (0)$ is the sum of contributions from low-energy fermions in all six patches. For $\Pi (Q)$ and $\Pi (Q+2k_F)$ such contributions come from two patches, and for $\Pi (2k_F)$ the low-energy contribution comes from only one patch. Away from an immediate vicinity of van Hove filling, all individual low-energy contributions are roughly of the same order.
Then $\Pi (0)$ term is the largest (because of the largest number of contributions), i.e., the $U^2_0$ term in $\lambda_f$ is attractive and can easily
 exceed a small repulsive bare term $\lambda^{(1)}_{t,II} \approx -4 U_1$ (see Eq. \ref{sa_9}).  As the consequence, $f-$wave channel remains attractive.  However, right at van Hove filling this is no longer the case because $\Pi (2k_F)$ diverges logarithmically, as $\log{\Lambda/T}$,  and
 overshoots $\Pi (0)$. The terms  $\Pi (Q)$ and $\Pi (Q+2k_F)$ diverge even stronger, as $\log^2 (\Lambda/T)$, but the two are identical at $\mu =0$ and cancel out in Eq. (\ref{sa_18}).  As the result, the $U^2_0$ contribution to  $f-$wave coupling in (\ref{sa_18}) is negative at $\mu =0$, i.e., $f-$wave channel becomes repulsive.
  In Fig.\ref{fig:13} we combine the results for $\lambda_f$ near van Hove and near full filling  and sketch the behavior of $\lambda_f$ at fermionic densities  between the two limits.  We clearly see that $\lambda_f$ is non-monotonic and has a maximum somewhere between van Hove and full filling.
  The non-monotonic behavior of the coupling combined with the $t-\mu$ dependence of the prefactor for $T_c$
   gives rise to a non-monotonic behavior of the onset temperature  for $f-$wave pairing, like in Fig.\ref{fig:13}

  We next analyzed patch model I, in which, we remind, the gap is concentrated near van-Hove points.  We found that the most relevant KL contribution near van Hove filling is in the doubly degenerate (symmetric) $d-$wave channel. The
   coupling in this channel is
   \beq
   \lambda_f = \lambda^{(2,s)}_{s,I} + \frac{U^2_0}{2} \left[-\Pi (0) - \Pi (2k_F) + \Pi (Q) + \Pi (Q + 2k_F)\right]
\label{sa_19}
\eeq
and, we remind, at small $\mu$,
 $\lambda^{(2,s)}_{s,I} \approx -U_1$.  The terms $\Pi (Q)$ and  $\Pi (Q + 2k_F)$ are identical at $\mu =0$ and both diverge as
 $\log^2 {\Lambda/T}$, hence at small $T$ the $U^2_0$ term definitely exceeds the bare repulsion. Then $\lambda_f  >0$, i.e., $d-$wave channel is attractive.

 Like we already said, the presence of the logarithms implies that one
 cannot restrict with the lowest order in perturbation theory and have
 to sum infinite series of KL-type  corrections to the pairing
 interaction. We discuss this in detail in Sec. 4.  The conclusion is
 qualitatively the same as the one from the present
 consideration.  Namely,
  the two $d-$wave
   components are degenerate and attractive near van-Hove
 filling, and
 the corresponding $T_c$  has a maximum value right at van-Hove filling
  (see e.g. Refs.~\onlinecite{Gonzalez,ChiralSC,Ronny1}.  The superposition of the two
 degenerate states can be chosen such that they are related by time-reversal. Below $T_c$,
  the system spontaneously chooses  one of the states in order to maximize
  condensation energy and, by doing so, breaks time-reversal symmetry.

 We combine the results for $\lambda_d$ at small $\mu$ and at $\mu \approx t$ and sketch the behavior of $d-$wave coupling between van Hove and full filling
  in Fig. \ref{fig:13}b.  The comparison between this  figure and Fig. \ref{fig:13}a shows that the system undergoes a transition from $d-$wave to an $f-$wave pairing at some distance from van Hove doping (Fig.\ref{fig:13}c).  At a truly weak coupling, $T_c$ is the largest right at van-Hove filling. 
   both attractive coincide is beyond the accuracy of our approach. It is also quite possible that SDW order sets in first in at least some range of dopings.

Like we said, we analyzed other channels, i.e., symmetric and antisymmetric $p-$wave, symmetric $f-$wave and antisymmetric $d-$wave, and found that they are not competitors to
symmetric $d$ and antisymmetric $f$-channels.  We also analyzed the pairing at smaller fillings, when $\mu <0$. We found that $f-$wave channel again is the most attractive,
 because $\Pi (0)$ is the largest.  Hence, as filling decreases from the full one, the largest attractive coupling evolves from $f-$wave to $d-$wave, and then again to $f-$wave.  This agrees with the numerical study in Refs.~\cite{scal_kiv_raghu,cho}.  We note that $f-$wave gap for $\mu <0$, where the FS consists of one piece, changes sign six times as one moves along the FS.

 \section{fermions on a honeycomb lattice}
\label{sec:3}

 A similar but not identical superconductivity emerges in KL analysis of interacting system of fermions on a honeycomb lattice.  This is what we analyze next.

\subsection{FS and fermionic dispersion}

\begin{figure}
 \includegraphics[width = \columnwidth]{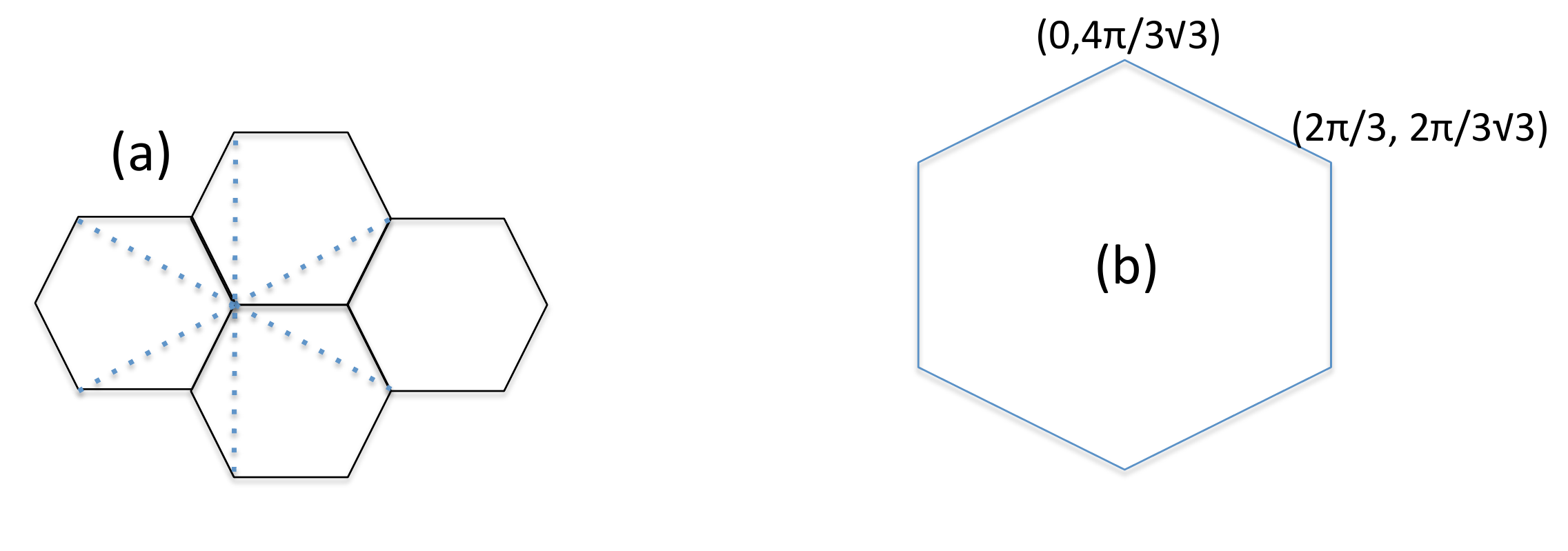}
 \caption{\label{fig:14}
 A honeycomb  lattice (a) and the corresponding Brillouin zone in momentum space (b).}
 \end{figure}
A honeycomb lattice is presented in Fig.\ref{fig:14}.
  There are two non-equivalent lattice sides, marked $K$ and $K'$ in the figure.
In a tight-binging model, fermion hops from one sublattice to the other. The corresponding Hamiltonian in a momentum space is~\cite{wallace}
\beq
{\cal H}_2 = - \sum_k \left[\gamma_k a^{\dagger}_k b_k +  \gamma^{*}_k b^{\dagger}_k a_k  -\mu \left(a^{\dagger}_k a_k + b^{\dagger}_k b_k\right)\right]
\eeq
 where
 \beq
 \gamma_k = t \left(e^{-ik_x} + 2 e^{ik_x/2} \cos{\frac{k_y \sqrt{3}}{2}}\right)
 \eeq
 It is convenient to express a complex $\gamma_k$ as
 $\gamma_k =  \epsilon_k e^{i \phi_k}$, where
 \beq
 \epsilon_k = |\gamma_k| = t \sqrt{1 + 4 \cos{\frac{3 k_x}{2}} \cos{\frac{k_y \sqrt{3}}{2}} + 4 \cos^2{\frac{k_y \sqrt{3}}{2}}}
 \label{sa_20}
 \eeq
  and
  \beq
 \cos{\phi_k} = \frac{\cos{k_x} + 2 \cos{\frac{k_x}{2}} \cos{\frac{k_y \sqrt{3}}{2}}}{\epsilon_k}, ~~ \phi_{-k} = - \phi_k.
 \label{sa_21}
 \eeq

 The quadratic form is diagonalized by  unitary transformation
 \bea
 a_k &=& \frac{1}{\sqrt{2}} e^{i\phi_k/2} \left( c_k + d_k\right), \nonumber \\
 b_k &=& \frac{1}{\sqrt{2}} e^{-i\phi_k/2} \left( c_k - d_k\right),
\eea
The excitation spectrum consists of two branches with opposite signs of energy
\beq
{\cal H}_2 = \sum_k \left[(\epsilon_k-\mu) d^{\dagger}_k d_k - (\epsilon_k + \mu) c^{\dagger}_k c_k \right]
\label{sa_22}
\eeq
At half-filling, $\mu =0$, FS consists of 6 Dirac points $(\pm 2\pi/3, \pm 2\pi/(3\sqrt{3})$, $(0, \pm 4\pi/(3\sqrt{3})$ at which $\epsilon_k =0$. Only two of these six Dirac points are inequivalent (the rest are connected by reciprocal lattice vectors), but it is convenient for our analysis to work keep track of excitations near all six Dirac points. The same results are (of course) obtained if we identify points that differ by a reciprocal lattice vector.
When filling either increases or decreases, 6 disconnected segments of the FS open up.  At $3/8$ or $5/8$ filling, when $|\mu| = t$, disconnected FS pieces merge.
 At these two fillings, the excitation spectrum has six van Hove points at $(\pm 2\pi/3, 0)$ and $(\pm \pi/3, \pm \pi/\sqrt{3})$ and  the FS obtained from (\ref{sa_20})
  consists of straight lines connecting van Hove points.  The van Hove points are protected by symmetry and survive when hopping extends beyond nearest neighbors.
  The nesting (FS in the form of straight lines) survives when the hopping extends to second neighbors, but gets destroyed by third and further neighbor
  hoppings.  At larger $|\mu|$, the FS has one piece, centered in the middle of the BZ. The size of the FS shrinks and it vanishes at $|\mu| = 3t$.
This evolution is quite similar to the one in Fig.\ref{fig:5} for the triangular lattice.

  For positive $\mu$ (filling above $1/2$), the six disconnected FSs are of  electron type (the states inside the FS have smaller energy than $\mu$) and
   the FS centered at $(0,0)$ is of hole type.  For negative $\mu$ the situation is opposite. The cases of positive and negative $\mu$ give identical results for superconductivity, and for definiteness below we focus on $\mu >0$ when $d_k$ fermions have a FS (see Eq. (\ref{sa_22})).

\subsection{The pairing interaction}

  We use the same general reasoning  as in previous paragraph. Namely, we split the antisymmetrized interaction into spin-singlet and spin-triplet components
    and further split  spin-singlet component into $s-$wave
    ($\lambda^{(1)}_s$) and doubly degenerate $d-$wave
    ($\lambda^{(2)}_s = \lambda^{(3)}_s$) and split spin-triplet
    component into $f-$wave ($\lambda^{(1)}_t$) and doubly degenerate
    $p-$wave ($\lambda^{(2)}_t = \lambda^{(3)}_t$).  By analogy with
    the triangular lattice case, we expect the most interesting
    physics of the pairing to develop in the  range of fillings between $1/2$ and $5/8$, when the FS consists of six disconnected segments. To shorten the presentation and not to bother a reader with the re-derivation of the same results as in the previous,
     we only consider the range of doping where the FS consists of six disconnected segments. We consider patch model I, in which the pairing predominantly involves
      fermions from momenta near where van-Hoe points are located at van-Hove doping. Like in the triangular case, we consider two types of the solutions for the gap: the "symmetric" solution,
     in which the gap remains of the same sign at the two ends when we split each van Hove point into two, and the "antisymmetric" one, in which
 the gap changes sign between the two split van-Hove points.  In another parallel to the triangular case, we also consider patch model II, in which we focus on the middle points of the disconnected FS segments.  We again find that the solutions for the gap in the patch model-II are similar to the solutions of the antisymmetric model I in $d-$wave and $f-$wave channels and to the solutions of the symmetric model I in
  $p-$wave and $s-$wave channels.

  The structure of the gap in various channels in these two patch models is quite similar to the one in triangular lattice (see Figs. \ref{fig:10} - \ref{fig:E2}), the only difference is that the FS for the honeycomb lattice is rotated by $90^o$ compared to that in the triangular lattice.  By this reason, we will not present separate figures for the gap structures  on a honeycomb lattice.

\subsection{First order in $U(q)$}

 Like in the triangular lattice case, we consider the lattice model with short-range interaction which extends up to second neighbors.
The interaction is originally written in real space in terms of lattice operators $a$and $b$. To re-express it in terms of operators $d_k$ and $d^\dagger_k$
 one needs to apply the unitary transformation, i.e., include the phase factors.  These phase factors are often neglected in the literature. We will see, however,
  that they play a certain role in our case.

\subsubsection{ Extended Hubbard model}

 We consider the model with interaction up to second neighbors.
 The on-site interaction potential involves fermions from the same sublattice and  phase factors cancel out for incoming momenta $k, -k$  and outgoing momenta $p, -p$:
 \begin{widetext}
 \beq
 H^{(0)}_{int} = -\frac{U_0}{2} \left(a^\dagger_{k,\alpha} a^\dagger_{-k,\beta} a_{p,\alpha} a_{-p,\beta} + b^\dagger_{k,\alpha} b^\dagger_{-k,\beta} b_{p,\alpha} b_{-p,\beta}\right) =  -\frac{U_0}{4} d^\dagger_{k,\alpha} d^\dagger_{-k,\beta} d_{p,\alpha} d_{-p,\beta}
\label{sa_23}
\eeq
\end{widetext}
Interaction between nearest neighbors involves fermions from different sublattices, and  phase factors  do not cancel and have to be kept~\cite{hexa-review}.  We obtain
 \beq
 H^{(1)}_{int} = -\frac{U_1}{8} d^\dagger_{k,\alpha} d^\dagger_{-k,\beta} d_{p,\alpha} d_{-p,\beta}   \epsilon_{k-p} \cos(\phi_{k-p}-\phi_k+\phi_p)
\label{sa_23_1}
\eeq
Interaction between second neighbors again involves fermions from the same sublattice and phase factors cancel out:
\bea
 &&H^{(2)}_{int} = -\frac{U_2}{2} d^\dagger_{k,\alpha} d^\dagger_{-k,\beta} d_{p,\alpha} d_{-p,\beta} \Phi (k-p) \nonumber \\
 && \Phi (q) = \cos{q_y \sqrt{3}} + 2 \cos{\frac{3 q_x}{2}} \cos{\frac{q_y \sqrt{3}}{2}}
 \label{sa_24}
\eea
Like before, we assume $U_0 >> U_1 >> U_2$ and will only keep the leading term in the formulas for the eigenvectors.\\

{\it{1. Van-Hove filling, $\mu =t$}}\\

Consider first the patch model I, which is centered at van Hove points. For a given triad $k_1 = (-2\pi/3, 0)$,  $k_2 = (\pi/3, \pi/\sqrt{3})$ and
$k_3 = (\pi/3, -\pi/\sqrt{3})$ we have $\phi_{k_i} = -\pi/3$ ($i=1,2,3$) and $\phi_{k_i-k_j} = \pi$ for $i \neq j$. Like in triangular case, there are only two
 different interactions $U(0) = U(2k_F)$ and $U(Q) = U(Q + 2k_F)$ where $Q= k_1-k_2$. A simple calculation yields
\beq
U(0) = \frac{U_0}{2} + \frac{3 U_1}{4} + \frac{3 U_2}{2}, ~ U(Q) = \frac{U_0}{2} - \frac{U_1}{4} - \frac{U_2}{2}
\label{sa_25}
\eeq
The interaction in the spin-triplet channel then vanishes identically, and in spin-singlet channel we obtain
\beq
  \lambda^{(1)}_s \approx -3\frac{U_0}{2} -U_1 - U_2,
  ~~  \lambda^{(2)}_s = \lambda^{(3)}_s = - U_1 - U_2
  \label{sa_26}
  \eeq
Both couplings are repulsive if $U_i$ are all positive.

The patch model II at van Hove filling is centered at $\pm {\bar k}_1$, $\pm {\bar k}_1$, and $\pm {\bar k}_1$, where
$ {\bar k}_1 = (0, \pi/\sqrt{3})$, ${\bar k}_2 = (-\pi/2, -\pi/(2 \sqrt{3})$, and ${\bar k}_3 = (\pi/2, -\pi/(2 \sqrt{3})$. For this model,
 the eigenvalues are non-zero in all channels. We obtained
   \bea
  &&\lambda^{(1)}_{s,II} = -3\frac{U_0}{2},   \lambda^{(2)}_{s,II} =  \lambda^{(3)}_{s,II} = -\frac{U_1}{4} \nonumber \\
   &&\lambda^{(1)}_{t,II} = - 2 U_2 ,~~  \lambda^{(2)}_{t,II} =  \lambda^{(3)}_{t,II} =
   - \frac{3U_1}{4}
    \label{sa_27}
   \eea
 Again, all eigenvalues are negative (repulsive),  when all $U_i >0$.\\

{\it{2. Larger filling, $\mu > t$}}\\

The analysis parallels the one for fermions on a triangular lattice and we just list the results.
The patch model I is centered at $(\pm \frac{2 (\pi -\delta)}{3},0)$ and $(\pm \frac{(\pi -\delta)}{3},\pm \frac{(\pi -\delta)}{\sqrt{3}})$,
where $\delta$ is related to $\mu$ as $\mu = t \sqrt{1+ 8 \sin^2 {\delta/2}}$. The center for each patch is the saddle point in the fermionic dispersion, e.g., near
$ k_1 = (-\frac{2 (\pi -\delta)}{3},0)$,
\bea
&&\epsilon_{k_1 + k} = -3 (\sin \delta/\sqrt{1 + 8 \sin^2 {\delta/2}}) k_x \nonumber \\
&& - (3/2) ((1 - 0.5 \cos \delta)/\sqrt{1 + 8 \sin^2 {\delta/2}}) k^2_y.
\eea
The eigenvalues are
  \bea
  &&\lambda^{(1)}_{s} = -3\frac{U_0}{2},   \lambda^{(2)}_{s} =  \lambda^{(3)}_{s} = -\frac{U_1}{4} \frac{(1 + \cos \delta)^2}{5 -4 \cos{\delta}} \nonumber \\
  &&  \lambda^{(1)}_{t} = 0,~~  \lambda^{(2)}_{t} =  \lambda^{(3)}_{t} =
   - \frac{9U_1}{4}\frac{\sin^2{\delta}}{5 -4 \cos{\delta}}
    \label{sa_28}
   \eea
 Observe that the coupling in $f-$wave channel vanishes, i.e., individual contributions from $U_0$, $U_1$ and $U_2$ all cancel out.

 The patch model II is centered at $(0, \pm \frac{(\pi +{\bar \delta})}{\sqrt{3}})$ and $(\pm \frac{(\pi +{\bar \delta})}{2},\pm \frac{(\pi +{\bar \delta})}{2\sqrt{3}})$. where ${\bar \delta} = - 2 \arcsin{(\mu-t)/2}$.
  The eigenvalues are
  \begin{widetext}
  \begin{equation}
  \lambda^{(1)}_{s,II} = -3\frac{U_0}{2},   \lambda^{(2)}_{s} =  \lambda^{(3)}_{s} = -\frac{U_1}{4} (1 + \sin{{\bar \delta}/2})^2
   \qquad \lambda^{(1)}_{t,II} = -2 U_2 \cos^2{{\bar \delta}/2} (1 +\sin{{\bar \delta}/2})^2,   \lambda^{(2)}_{t,II} =  \lambda^{(3)}_{t,II} =
   - \frac{2U_1}{8} (1 + \cos {{\bar \delta}})
    \label{sa_29}
   \end{equation}
   \end{widetext}
All bare couplings are again repulsive. Note, however, that  the coupling in f-wave channel is proportional to second-neighbor interaction potential
$U_2$, i.e., it is smaller than the repulsive couplings in other channels.\\

{\it {3. Smaller fillings $0 <\mu < t$}}\\

In this range of fillings, the FS consists of 6 segments around Dirac points.
For the patch model II the eigenfunctions are the same as in Eq. (\ref{sa_29}), only
 now ${\bar \delta}$ is positive and varies between ${\bar \delta} =0$ at $\mu =t$ and ${\bar \delta} = \pi/3$ at $\mu =0$.
 The couplings are all repulsive, the smallest is in $f-$channel, and this one varies between $\lambda^{(1)}_{t,II} =-2U_2$ at $\mu =0$ and $
 \lambda^{(1)}_{t,II} = -27 U_2/8$ at $\mu = t$.

 Like in the triangular lattice case, the solutions for the gap for the patch model I
 split into symmetric and antisymmetric subclasses, which differ in the relative signs of the gaps at split van Hove points ($k_{i+}$ and $k_{i-}$). The formulas for
 the couplings are rather complicated due to the presence of coherence factor in nearest-neighbor interaction and we do not present them.
  The key result is that the couplings are repulsive in all channels except in
  symmetric spin-triplet channels and  antisymmetric spin-singlet channels, where
   individual contributions of order  $U_1$ and $U_2$ cancel out  for all $0<\mu <t$.
   In technical terms, the cancelation holds between $U(0)$ and $U(k_{1+} + k_{1-})$, between $U(k_{1+}-k_{2+})$ and $U(k_{1+} + k_{2,-})$, etc.

To summarize, we see that the couplings to order $U(q)$ are generally repulsive, although the one for the $f-$wave channel in the model II  (similar to the one in the antisymmetric $f-$wave channel in model I) is only repulsive due to second-neighbor $U_2 >0$.  As a peculiarity of the honeycomb lattice, the coupling in
the  symmetric spin-triplet $f-$wave channel vanishes for all dopings, and the couplings in
the symmetric $p-$wave channel and antisymmetric $s$ and $d-$channels  vanish for  all $0<\mu <t$. The antisymmetric solution in the patch model I indeed only makes sense when the FS consists of separate segments.

\subsection{Kohn-Luttinger renormalizations}

We now discuss the KL physics -- the renormalization of the interactions to second order in $U$. Like in the triangular case, we search for the renormalizations
  coming from on-site interaction as attraction at order $U^2_0$ overshoots bare repulsion of order $U_1$ or, even better, $U_2$, already at weak coupling.

We again consider the two limiting cases  -- one  close to half-filling, when FS segments around Dirac points are small, and the other near van Hove filling.

\subsubsection{Near half filling}

We first consider patch model I and look into the channels for which the couplings to order $O(U)$  vanish.  These are antisymmetric spin singlet and symmetric spin-triplet channels.  If there was an attraction in any of these channels from KL terms, the system would become a superconductor for arbitrary weak interaction $U(q)$. We found, however, that this does not happen. Namely, the KL renormalizations in these particular channels involve the combinations of polarization operators
\begin{widetext}
\beq
\Pi (0) - \Pi (k_{1+} + k_{1-}), ~ \Pi (k_{1+} - k_{1-}) - \Pi (2 k_{1+}),~\Pi (k_{1+} - k_{2+}) - \Pi (k_{1+} + k_{2-}),~\Pi (k_{1+} - k_{2-}) - \Pi (k_{1+}+ k_{2+})
\label{sa_30}
\eeq
\end{widetext}
  A straightforward analysis shows that these combinations are all zero. For example, for $k_{1+} = (2\pi/3, \delta/\sqrt{3})$, $k_{1-} = (2\pi/3, -\delta/\sqrt{3})$,
  where $\delta = 2 \arccos{(t-\mu)/2t}$, $\epsilon_{k_{1+} + k_{1-} +q} = \epsilon_q$ and $\epsilon_{k_{1+} - k_{1-} +q} = \epsilon_{2 k_{1+} +q}$.  Elementary analysis then shows up that each of the terms in Eq. (\ref{sa_30}) vanishes.   The implication is that, in patch model I, the  pairing interaction in these particular channels vanishes to order $U^2$ and, very likely, to all orders in $U$. This vanishing is the consequence of FS nesting, which is present  as long as one restricts with the hopping between nearest and second nearest neighbors.  Hopping between third  neighbors breaks nesting and gives rise to  non-zero couplings in antisymmetric spin singlet and symmetric spin-triplet channels.  The third neighbor hopping is, however rather weak, at least in graphene, and these pairing interactions, even if attractive,  are
   weak compared to interactions in antisymmetric $f-$wave channel and symmetric $d-$wave channel, which we consider below.

An antisymmetric $f-$wave and symmetric $d-$wave chanels can be conveniently studied at small $\mu$ within patch model II, and we now focus on this model.
Like in the case of a triangular lattice,  there are  two relevant interactions $U_{ren} (0)$ and $U_{ren} (2k_F)$.
Keeping only $U_0$ term we obtain, like before,
    \bea
&&U_{ren}(0) = \frac{U_0}{2} + \frac{1}{4} U^2_0 \Pi (2k_F), ~ U_{ren}(2k_F) = \frac{U_0}{2} + \frac{1}{4} U^2_0 \Pi (0) \nonumber \\
&&U_{ren}(Q) = \frac{U_0}{2} + \frac{1}{4} U^2_0 \Pi (2k_F), \nonumber \\
&& ~ U_{ren}(Q + 2k_F) = \frac{U_0}{2} + \frac{1}{4} U^2_0 \Pi (0)
\label{sa_31}
\eea
where $Q$ is approximately the distance between Dirac points.  Again, the polarization operator $\Pi (0)$ at zero momentum transfer is approximately two times larger than
 the one at $2k_F$ because $\Pi (0)$ collects the contributions from low-energy fermions from all six segments of the FS, while $\Pi (2k_F)$ collects low-energy contributions from three segments, and in other three the momentum transfer by $2k_F$  places a fermion far away from the FS.  As a result, the coupling in the $f-$wave channel is
      \bea
\lambda_f &=& \lambda^{(1)}_{t,II}  + \frac{3}{8} U^2_0 \left(\Pi(0) - \Pi (2k_F)\right) \nonumber \\
&& \approx -27 \frac{U_2}{8}
  + \frac{3}{16} U^2_0 \Pi(0)
    \label{sa_32}
   \eea
 We see that, like in the triangular case, the $U^2_0$ term in the interaction is attractive ($\Pi (0)$ is positively defined).  Because $U_0 >> U_2$, the KL term overshoots the bare repulsion already at weak coupling. Once this happens, the system becomes unstable against $f-$wave pairing of the same type as Fig.\ref{fig:E2}.

There is one difference with the triangular case though.
The evaluation of $\Pi (0)$ yields~\cite{hwang}
 \beq
 \Pi (0) = \frac{2 \mu}{\sqrt{3}\pi t^2} + O(\mu^2)
 \label{sa_33}
 \eeq
 In distinction to the case of a triangular lattice, the polarization operator $\Pi (0)$ scales linearly with $\mu$ and vanishes right at half-filling.
 This implies that $f-$wave attraction develops only at some small but finite distance away from from half-filling, when the KL attractions develops enough to overshoot the first-order repulsion.  An additional smallness at $\mu << t$ comes from the
  fact that the actual parameter which appears in the exponent in the BCS formula for $T_c$ is the product of $\lambda^{(1,ren)}_{t,II}$ and the density of states at the FS. The latter also scales with $\mu$:
  \beq
  N_F = \frac{\mu}{3\pi t^2} + O(\mu^2)
  \label{sa_34}
  \eeq
  Introducing the dimensionless coupling ${\bar \lambda} = \lambda N_F$, we obtain  for $f-$wave channel
  \beq
  {\bar \lambda}_f =  - \frac{9}{8\pi} \left(\frac{U_2}{t}\right)^2 \left(\frac{\mu}{t}\right) +
  \frac{\sqrt{3}}{24\pi^2} \left(\frac{U_0}{t}\right)^2 \left(\frac{\mu}{t}\right)^2
 \label{sa_35}
 \eeq
 Superconducting $T_c \sim \mu e^{-1/{\bar \lambda}_f}$.  At small $\mu/t$, $T_c$ is vanishingly small, if not zero.
   However at, say, $U_0 \sim 6t$ (the bandwidth) and $\mu \leq  t$
    the dimensionless coupling is ${\bar \lambda}_f \approx 0.26$ and $T_c \sim 10^{-3} t$ which is not small given that $t \sim 10^4 K$.

\begin{figure}
 \includegraphics[width = \columnwidth]{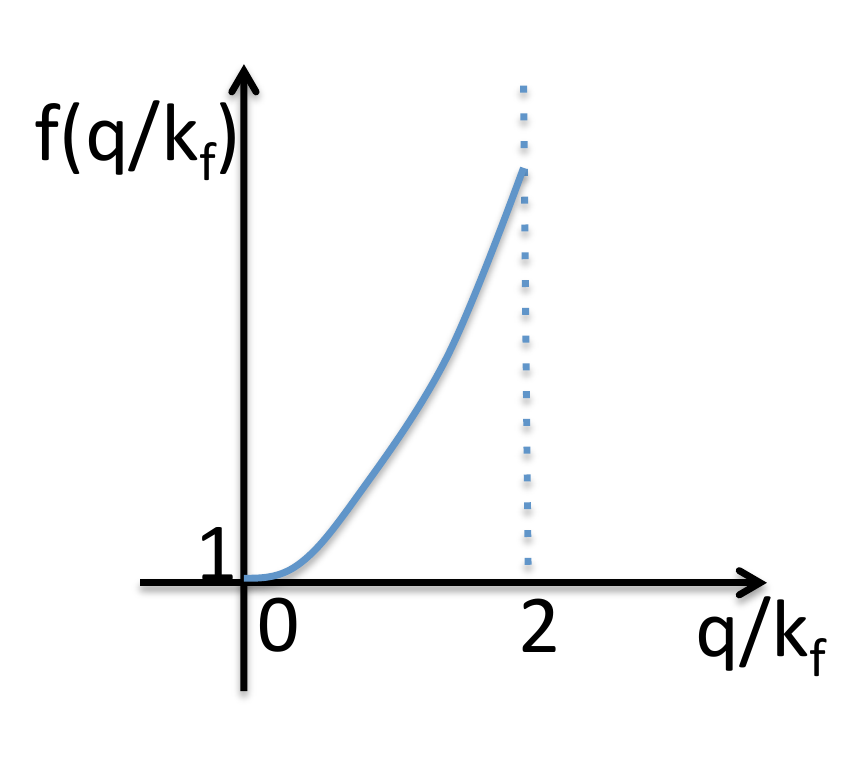}
 \caption{\label{fig:18}
 Scaling function $f\left(q/k_F\right)$ in Eq. \protect\ref{sa_36}. The limiting values are $f(0) =1$, $f(2) = 3.66/\pi$.}
 \end{figure}

There is an interesting peculiarity of $f-$wave pairing on a honeycomb lattice -- the gap has a non-vanishing angular dependence at small $\mu$
 This follows from the fact that if one calculate $\Pi (q)$ for Dirac fermions at small $\mu$ without approximating $\Pi (q)$ by $\Pi (0)$, we obtain
  universal dependence on $q/k_F$:
  \beq
  \Pi (q) = \frac{2\mu}{\pi \sqrt{3} t^2} f\left(\frac{q}{k_F}\right)
  \label{sa_36}
  \eeq
   where $k_F = 2\mu/(3t)$.  In the two limits $f(0) =1$ and $f(2) = 3.66/\pi$. In between,   $f(x)$ interpolates as shown in Fig.\ref{fig:18}.
   Because of this dependence, if we label by $\phi$ and $\phi'$ the angles specifying the locations of fermions on the FS's ($-\pi/3 < \phi, \phi' < \pi/3$),
    the kernel in the gap equation relating $\Delta (\phi)$ and $\Delta (\phi')$ contains the function of $\cos (\phi - \phi')$.
    A symmetric in $\phi$ solution is then $\Delta (\phi) = \Delta (1 + \alpha \cos \phi + \beta \cos^2 \phi +...)$, where $\alpha$, $\beta$, etc are universal numbers independent on $\mu$.

\begin{figure}
 \includegraphics[width = \columnwidth]{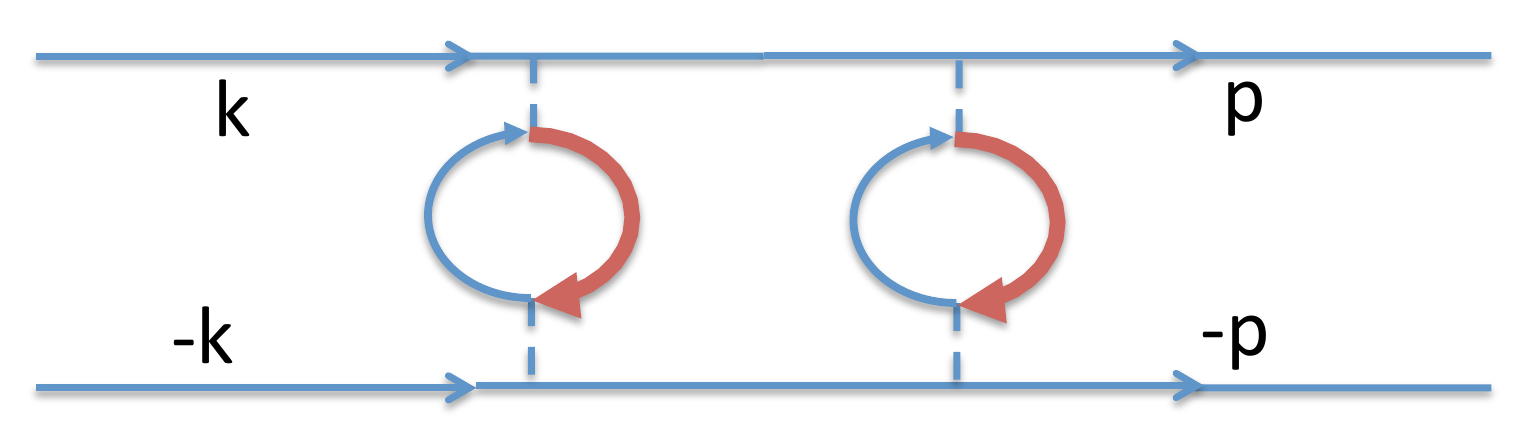}
 \caption{\label{fig:E3} An example of higher-order diagram which contributes to KL renormalization of the pairing interaction near half-filling.
 Thin blue and thick red lines represent fermions from the two branches with positive and negative energies. The dashed line is the Hubbard interaction $U(0)$.}
 \end{figure}

The word of caution. At second order in $U$, the presence of the other branch of excitations ($c-$ branch at negative energies in Eq. (\ref{sa_22})) does not affect KL consideration by two reasons. First, the
 interactions in $d-c$ basis do not contain terms with three $d$-operators and one $c-$operator, hence there are no KL contribution with the bubbles made out of one $d-$ and one$c-$ fermion.  Second, the terms with two $c-$and two $d-$fermions are present, but these contain
the polarization bubble made of two $c-$ fermions. The latter  vanishes because the bubble is identically zero for any positive $\mu$.
  However, at higher orders, the diagrams of the type shown in Fig.\ref{fig:E3} do contribute.  These diagrams contain pairs of  polarization bubbles, each
   made out of one $c$ and one $d$-fermion.  Such a bubble tends to a finite value at vanishing $\mu$, hence the dimensionless coupling from such term contains only one power of $\mu$ (but higher power of $U/t$).  Whether these terms are friends or foes of $f-$wave superconductivity is unclear.

\subsubsection{Near van Hove points}

\begin{figure}
 \includegraphics[width = \columnwidth]{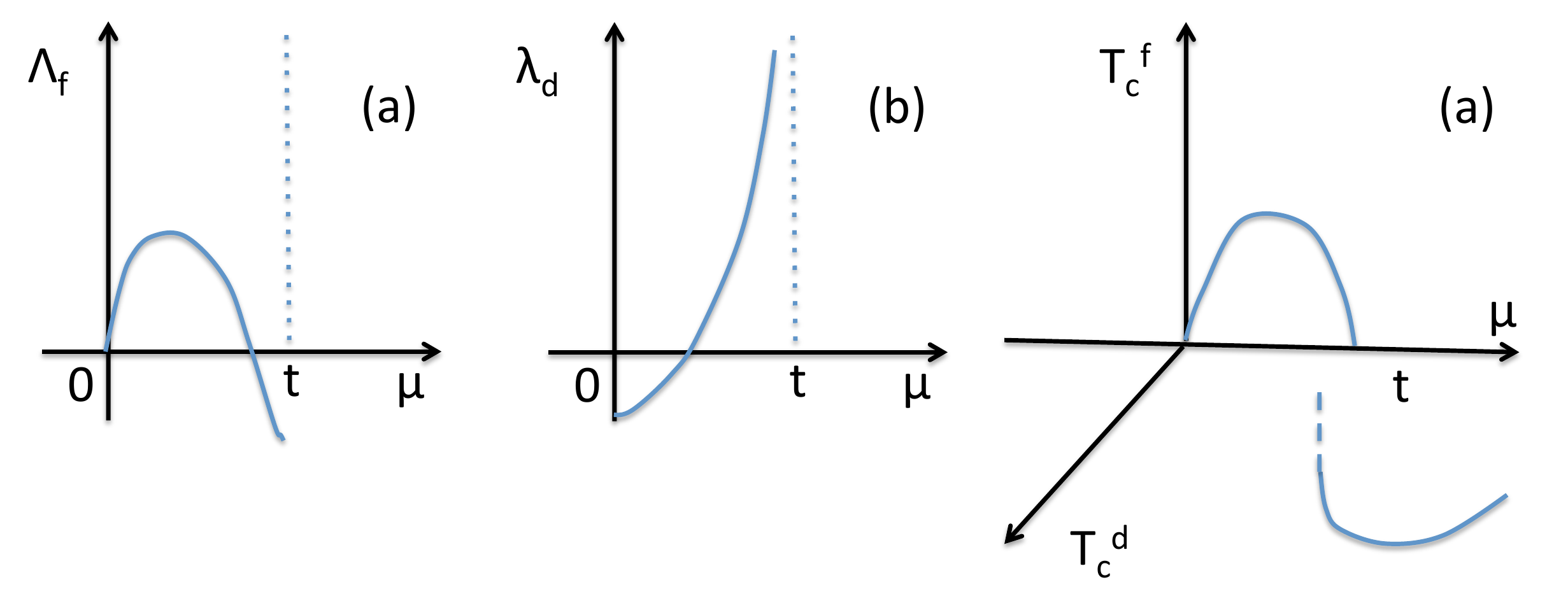}
 \caption{\label{fig:19} Fermions on a honeycomb lattice.  Panels  (a) and (b) -- the behavior of the leading eigenfunctions in $f-$wave and $d-$wave channels as functions of the chemical potential for $0 <\mu <t$, when the FS consists of six disconnected segments.  Panel (c) shows the behavior of $T_c$ in the two channels. The d-wave channel has the strongest instability (largest $T_c$) near Van Hove doping.}
 \end{figure}

The consideration near van Hove points is essentially identical to the one for fermions on a triangular lattice and we only state the result:
 to order $U^2_0$ interaction in symmetric $d-$wave channel in patch model I is attractive and is logarithmically singular.
 The coupling in $f-$wave channel in patch model II) is repulsive at this density.  The behavior of $\lambda_d$ and $\lambda_f$ at various densities is shown in Fig.\ref{fig:19} together with the corresponding $T_c$.

  As the consequence, within KL approximation, the strongest pairing instability at and near  van Hove points  is doubly generate $d$-wave.
  Whether there is a direct transition between $f-$wave at smaller $\mu$ and $d-$wave at $\mu \approx t$ (and possible co-existence of the two superconducting orders), or the system first looses $d-$wave order, remains normal down to $T=0$ in some range of $\mu \leq t$ and then becomes unstable towards $f-$wave
   superconductivity, depends on details. An SDW order at intermediate $\mu$ is also a possibility~\cite{Ronny1,dhlee_fawang,NandkishoreSDW,hexa-review}.
    Numerical study of the Hubbard model with only $U_0$ interaction indicates that superconductivity of one type or the other develops for all $\mu <t$.   How the situation changes due to interaction between further neighbors remains to be seen.

 At even larger $\mu >t$, the analysis of superconductivity becomes less universal because polarization operators $\Pi (0)$, $\Pi (2k_F)$, $\Pi (Q)$, and $\Pi (Q +2k_F)$ are all of the same order and the interplay between them depends on the details of the band structure.  If $\Pi (q)$ drops rather fast with increasing $q$, $f-$wave channel again becomes attractive and wins over other channels. If so, the system evolves from $f-$wave to $d-$wave and then again to $f-$wave superconductor as $\mu$ increases.  However, the coupling gets smaller as $\mu$ increases, so the best chance to detect $f-$wave superconductivity is to
  analyze the region $\mu \leq t$.

 In the rest of the paper we present more detailed analysis of superconductivity and its interplay with other instabilities near van Hove filling.
 Like we said, at van Hove doping the polarization operators at all four relevant momenta diverge logarithmically, either as $\log{\Lambda/T}$ or as $\log^2 {\Lambda/T}$,
 and  second-order KL analysis is not adequate.

\section{At the Van Hove point}
\label{sec:4}

When fermions on a triangular or honeycomb lattices with hopping between nearest neighbors
are doped to the saddle points of the dispersion, they
display a nested FS (FS) and a logarithmically divergent
density of states (DOS), which trigger instabilities to numerous strongly
ordered states. It has recently been shown using a RG
 calculation that the leading instability in the presence
of weak repulsive interactions is to chiral d wave superconductivity.
However, that calculation only took into account the `corners' of
the FS, which are saddle points of the dispersion and dominate the
DOS. Here, we present the RG for the full FS.

We divide the FS up into `corner' regions (which are close to a saddle
point and dominate the DOS) and `edge' regions. Ignoring the edges
altogether corresponds to the RG from
Ref. \cite{ChiralSC}. Here, we will take
the edge regions into account also. Initially, we work within a leading
log approximation (valid in the weak coupling limit), and demonstrate
that the behaviour of the FS edges mimics the behavior
of the corners, so that the full FS displays the same instability
as the corners. Thus, if the FS corners develop d wave superconductivity,
they induce d-wave superconductivity on the rest of the FS also.

Next, we extend the calculations away from the limit of weak coupling.
Guided by our earlier analysis of the weak coupling problem, we assume
that it is sufficient to determine the leading instability of the
FS corners, since the rest of the FS will be dragged along. We derive
the one loop RG equations for the corners, including correction terms
coming from the edge fermions. These corrections are subleading in
logs, however they must be taken into account when interactions are
not weak. We show that whereas at weak coupling there is only one
stable fixed trajectory, above a critical coupling strength a second
stable fixed trajectory appears. Thus, at weak coupling there is a
unique instability (corresponding to d-wave superconductivity), whereas
above a critical coupling strength there are two possible instabilities,
with the nature of the microscopic interactions determining which
one develops. The two instabilities are shown to correspond to d-wave
superconductivity and ferromagnetism respectively.

\subsection{The model}

For definiteness, in this Section we focus on a honeycomb lattice. Fermions on a triangular lattice show the same behavior.
We consider a system with nearest-neighbor hopping, described by the dispersion given by Eq. (\ref{sa_20}). We focus on Van Hove filling at $5/8$ density (a positive $\mu$) and consider only the band of low-energy fermions. We have
\begin{equation}
 \label{eq: dispersion}
\epsilon_{\vec{k}} = -t \sqrt{1 + 4 \cos{\frac{k_y \sqrt{3}}{2}} \cos{\frac{3 k_x}{2}} + 4 \cos^2{\frac{k_y \sqrt{3}}{2}}} - \mu
\end{equation}
The saddle point corresponds to setting the chemical potential $\mu=t.$ The FS takes the form of a hexagon inscribed within a hexagonal
BZ (Fig.) The corners of the hexagonal FS are saddle points
of the dispersion, and dominate the DOS.

We split up the FS into three inequivalent corners (labelled A,B and
C), and six inequivalent edges, labelled $\pm1,$$\pm2,$$\pm3$ as
shown in Fig.(fullFS). Crystal momentum conservation strongly restricts
the allows scattering processes. The allowed scattering processes
involving at least one fermion on edge 1 are enumerated in the table
below. The equivalent scattering processes involving fermions one
of the other edges can be deduced by symmetry. We also enumerate the
inequivalent corner-corner scattering processes involving corners
$A$ and $B$. The processes involving corner $C$ can again be deduced
by symmetry. Note that all these processes conserve momentum up to
a reciprocal lattice vector.

\begin{table*}
\begin{tabular}{|c|c|c|c|c|c|}
\hline
Corner-corner & Incoming state & Outgoing state & Edge-Edge & Incoming state & Outgoing state\tabularnewline
\hline
\hline
$g_{1}$ & $|(A,\sigma)$$,(B,\sigma')\rangle$ & $|(B,\sigma)$$,(A,\sigma')\rangle$ & $h_{1A}$ & $|(1,\sigma)$$,(2,\sigma')\rangle$ & $|(2,\sigma)$$,(1,\sigma')\rangle$\tabularnewline
\hline
$g_{2}$ & $|(A,\sigma)$$,(B,\sigma')\rangle$ & $|(A,\sigma)$$,(B,\sigma')\rangle$ & $h_{1A}$ & $|(1,\sigma)$$,(-3,\sigma')\rangle$ & $|(-3,\sigma)$$,(1,\sigma')\rangle$\tabularnewline
\hline
$g_{3}$ & $|(A,\sigma)$$,(A,\sigma')\rangle$ & $|(B,\sigma)$$,(B,\sigma')\rangle$ & $h_{1B}$ & $|(1,\sigma)$$,(3,\sigma')\rangle$ & $|(3,\sigma)$$,(1,\sigma')\rangle$\tabularnewline
\hline
$g_{4}$ & $|(A,\sigma)$$,(A,\sigma')\rangle$ & $|(A,\sigma)$$,(A,\sigma')\rangle$ & $h_{1B}$ & $|(1,\sigma)$$,(-2,\sigma')\rangle$ & $|(-2,\sigma)$$,(1,\sigma')\rangle$\tabularnewline
\hline
Corner-Edge & Incoming state & Outgoing state & $h_{1C}$ & $|(1,\sigma)$$,(-1,\sigma')\rangle$ & $|(-1,\sigma)$$,(1,\sigma')\rangle$\tabularnewline
\hline
$v_{1A}$ & $|(1,\sigma)$$,(B,\sigma')\rangle$ & $|(B,\sigma)$$,(1,\sigma')\rangle$ & $h_{2A}$ & $|(1,\sigma)$$,(2,\sigma')\rangle$ & $|(1,\sigma)$$,(2,\sigma')\rangle$\tabularnewline
\hline
$v_{1A}$ & $|(1,\sigma)$$,(C,\sigma')\rangle$ & $|(C,\sigma)$$,(1,\sigma')\rangle$ & $h_{2A}$ & $|(1,\sigma)$$,(-3,\sigma')\rangle$ & $|(1,\sigma)$$,(-3,\sigma')\rangle$\tabularnewline
\hline
$v_{1B}$ & $|(1,\sigma)$$,(A,\sigma')\rangle$ & $|(A,\sigma)$$,(1,\sigma')\rangle$ & $h_{2B}$ & $|(1,\sigma)$$,(-2,\sigma')\rangle$ & $|(1,\sigma)$$,(-2,\sigma')\rangle$\tabularnewline
\hline
$v_{2A}$ & $|(1,\sigma)$$,(B,\sigma')\rangle$ & $|(1,\sigma)$$,(B,\sigma')\rangle$ & $h_{2B}$ & $|(1,\sigma)$$,(3,\sigma')\rangle$ & $|(1,\sigma)$$,(3,\sigma')\rangle$\tabularnewline
\hline
$v_{2A}$ & $|(1,\sigma)$$,(C,\sigma')\rangle$ & $|(1,\sigma)$$,(C,\sigma')\rangle$ & $h_{2C}$ & $|(1,\sigma)$$,(-1,\sigma')\rangle$ & $|(1,\sigma)$$,(-1,\sigma')\rangle$\tabularnewline
\hline
$v_{2B}$ & $|(1,\sigma)$$,(A,\sigma')\rangle$ & $|(1,\sigma)$$,(A,\sigma')\rangle$ & $h_{3}$ & $|(1,\sigma)$$,(1,\sigma')\rangle$ & $|(-1,\sigma)$$,(-1,\sigma')\rangle$\tabularnewline
\hline
$v_{3A}$ & $|(1,\sigma)$$,(-1,\sigma')\rangle$ & $|(B,\sigma)$$,(B,\sigma')\rangle$ & $h_{4}$ & $|(1,\sigma)$$,(1,\sigma')\rangle$ & $|(1,\sigma)$$,(1,\sigma')\rangle$\tabularnewline
\hline
$v_{3A}$ & $|(1,\sigma)$$,(-1,\sigma')\rangle$ & $|(C,\sigma)$$,(C,\sigma')\rangle$ & $h_{5}$ & $|(1,\sigma)$$,(-1,\sigma')\rangle$ & $|(2,\sigma)$$,(-2,\sigma')\rangle$\tabularnewline
\hline
$v_{3B}$ & $|(1,\sigma)$$,(-1,\sigma')\rangle$ & $|(A,\sigma)$$,(A,\sigma')\rangle$ & $h_{5}$ & $|(1,\sigma)$$,(-1,\sigma')\rangle$ & $|(-3,\sigma)$$,(3,\sigma')\rangle$\tabularnewline
\hline
$v_{4}$ & $|(1,\sigma)$$,(B,\sigma')\rangle$ & $|(-1,\sigma)$$,(C,\sigma')\rangle$ & $h_{6}$ & $|(1,\sigma)$$,(-1,\sigma')\rangle$ & $|(-2,\sigma)$$,(2,\sigma')\rangle$\tabularnewline
\hline
$v_{4}$ & $|(1,\sigma)$$,(C,\sigma')\rangle$ & $|(-1,\sigma)$$,(B,\sigma')\rangle$ & $h_{6}$ & $|(1,\sigma)$$,(-1,\sigma')\rangle$ & $|(3,\sigma)$$,(-3,\sigma')\rangle$\tabularnewline
\hline
$v_{5}$ & $|(1,\sigma)$$,(B,\sigma')\rangle$ & $|(C,\sigma)$$,(-1,\sigma')\rangle$ &  &  & \tabularnewline
\hline
$v_{5}$ & $|(1,\sigma)$$,(C,\sigma')\rangle$ & $|(B,\sigma)$$,(-1,\sigma')\rangle$ &  &  & \tabularnewline
\hline
\end{tabular}
\caption{Table listing allowed scattering processes. Here $\sigma, \sigma'$ are spin labels, $A,B,C$ label inequivalent saddle points (FS corners), and $\pm1, \pm2, \pm3$ label FS edges. The allowed interactions are assumed to be short range (i.e. they are assumed to have no momentum dependence).}
\end{table*}

We now study the evolution of these various couplings under perturbative
RG. An RG framework is suitable for this problem, because the interactions
are marginal at tree level, with $\ln$ corrections in perturbation
theory. The presence of a logarithmic Van Hove singularity in the
density of states (DOS) means that the most divergent diagrams at
$n$ loop order in perturbation theory diverge as $\ln^{2n}\frac{{1}}{E}$
in the infrared. The perturbation theory breaks down once $g_{0}\ln^{2}\frac{{\Lambda}}{E}\sim O(1)$,
where $g_{0}$ is the bare coupling and $\Lambda$ is the ultraviolet
cutoff for the theory. In the weak coupling regime $g_{0}\ll1,$ we
have $\ln\Lambda/E$$\gg1$ at the limit of applicability of perturbation
theory. Therefore, diagrams subleading in powers of $\ln\Lambda/E$
can be ignored, and we can concentrate our attention on the most strongly
divergent diagrams. In this limit, the leading divergences can be re-summed into a $\ln^{2}$ RG.

The perturbation theory for the full FS contains `corner-corner' loops
(taken into account in the patch model), and also `edge-edge loops'
and `corner-edge' loops. We find that `edge-edge' and `corner-edge'
loops are at most $\ln^{1}$ divergent. Thus, they can be safely ignored
in the weak coupling limit, but they can introduce $\ln^{1}$ corrections
as we move away from weak coupling. Moreover, `corner-edge' loops
 are suppressed by a factor of $\Lambda/W$, where $W$ is the bandwidth ($W \sim 12 eV$ in graphene),
and we assume $\Lambda\ll t.$ We therefore neglect corner-edge loops throughout this paper. It thus follows that the $\ln^{2}$
RG is controlled by corner-corner loops, with $\ln^{1}$ corrections
coming from `corner-corner' and `edge-edge' loops.

We present the analysis in two sections: first, we calculate the full FS RG in the weak coupling limit. In this limit we can neglect subleading logs, and concentrate on the $\ln^2$ divergent diagrams, which come from corner-corner loops only. We will show that in this limit the only instability is to d-wave superconductivity. Next, we extend the analysis away from weak coupling.
by taking into account subleading $\ln^1$ corrections.
We will show that above a critical coupling strength there appears a
second instability, which corresponds to ferromagnetism. There is also
the possibility of a spin density wave instability (SDW), but this
requires us to stop the RG at some scale, since if the RG is allowed
to run indefinitely some other instability ultimately overtakes the
SDW.
 This  picture becomes even more diverse for the case of
strong nearest neighbor interactions~\cite{Raghu}.

\subsection{Full FS RG at weak coupling}

The building blocks of the RG are particle-particle ladders and particle-hole
bubbles, evaluated at momentum transfer zero and momentum transfer
equal to a nesting vector. The ladders and bubbles involving corner
fermions were presented in \cite{ChiralSC}. Here we present ladders and bubbles involving edge fermions.

For diagrams involving only edge fermions, the leading divergence
is logarithmic. There are no $\ln^{2}$ divergences because there
are no saddle points on the edge. The $\ln$ divergences in edge-edge
diagrams arise only in the Cooper and Pierls channels i.e. in the
particle-particle ladder at zero momentum, and in the particle-hole
bubble at momentum equal to one nesting vector. These take the form
\[
\Pi_{pp}^{edge}(0)=2c\xi;\qquad\Pi_{ph}^{edge}(Q)=2c\xi
\]
where and
$c$ is a cutoff dependent non-universal prefactor of order 1 and we have defined the RG scale $\xi=\frac12\nu_0 \ln\frac{\Lambda}{E}$, which is equal to one half the density of states at an energy $E$. The above expressions assume perfect nesting. Imperfect nesting may be dealt with by cutting off the growth of $\Pi_{ph}(Q)$ at some RG scale $\xi_c$, as discussed in \cite{ChiralSC}. All
other edge-edge diagrams are convergent. Meanwhile, it may be straightforwardly
verified that all ladders and bubbles constructed out of one corner
fermion and one edge fermion are suppressed by
 $\Lambda/W,$ where, we remind,
$W$ is the bandwidth
and $\Lambda$
is the UV cutoff for the RG. The supression arises because a corner-edge
pair can never be nested in any scattering channel, and thus the phase
space for these scattering processes is strongly restricted. We assume
$\Lambda/W \ll 1$, which allows us to neglect mixed corner-edge
loops. Moreover, in the weak coupling limit, $\ln^2$ divergent diagrams are much more important than $\ln$ divergent diagrams, which also allows us to neglect edge-edge loops. The resulting $\ln^2$ renormalizations come only from diagrams with corner-corner loops. The RG equations in the corner-corner sector are

\begin{figure}
a)\includegraphics[width = 0.45\columnwidth]{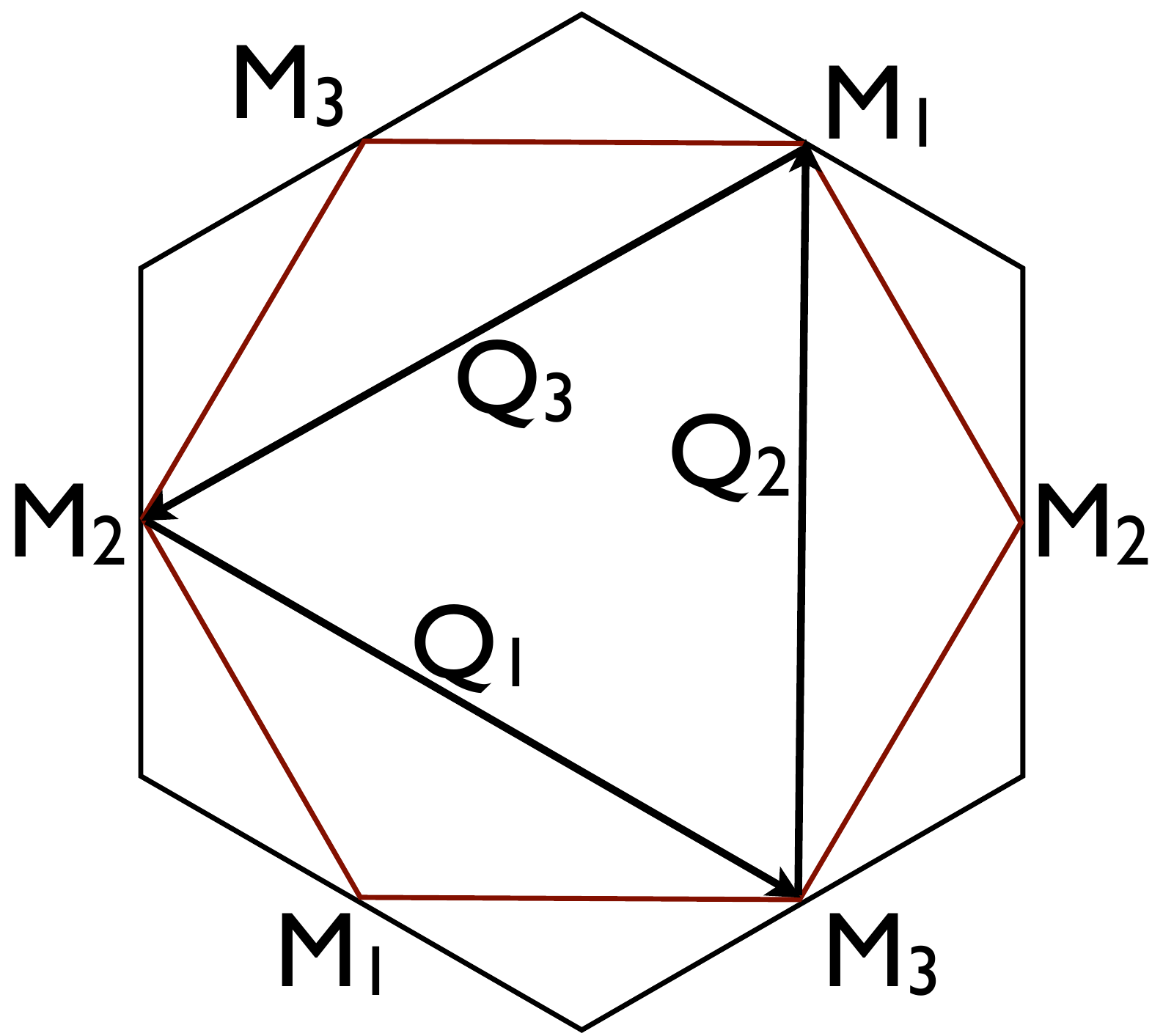}
b)\includegraphics[width = 0.45 \columnwidth]{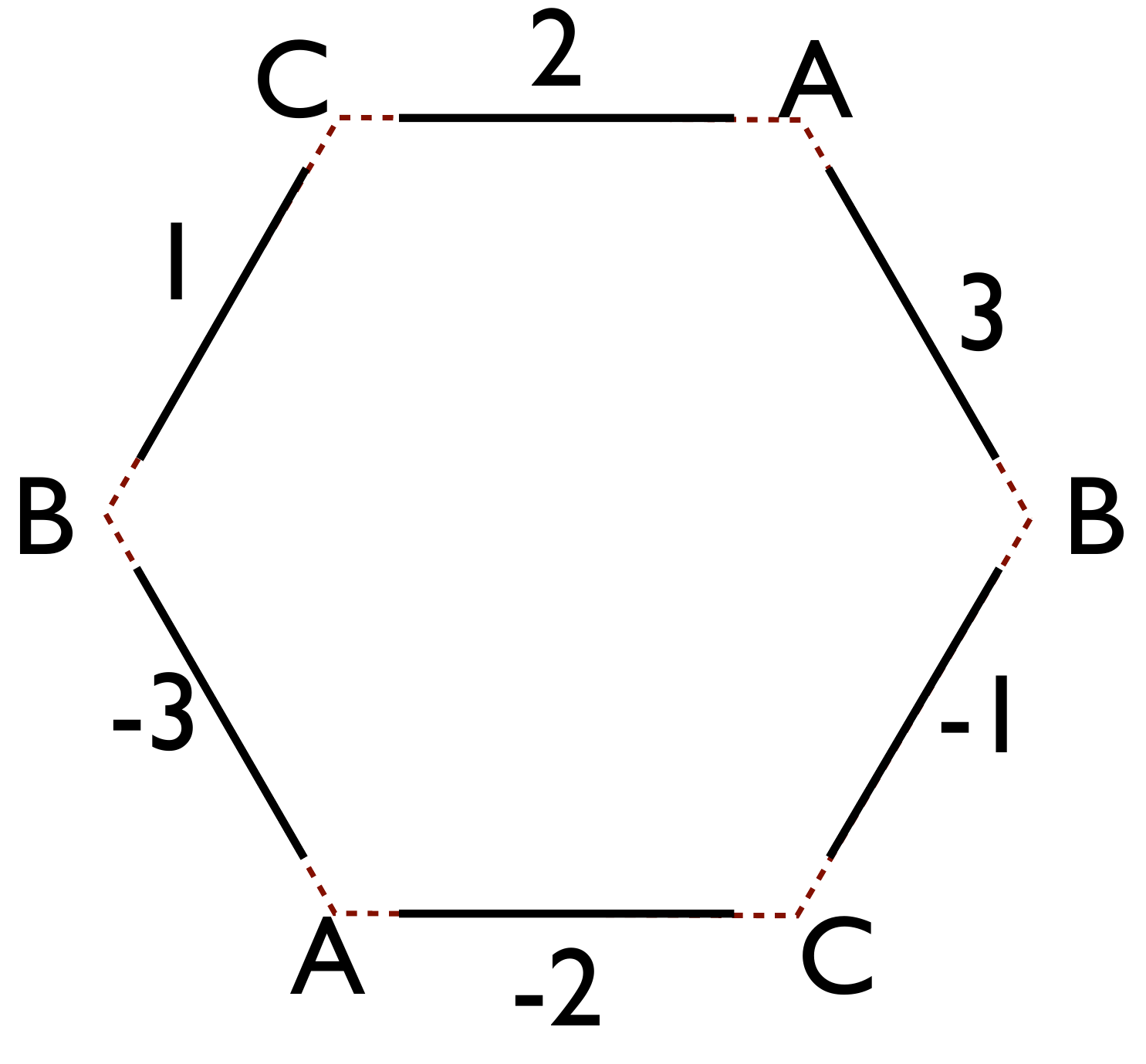}
\caption{a) Shows hexagonal FS inscribed within hexagonal BZ. The three inequivalent corners of the FS, $M_1, M_2, M_3$ are saddle points of the dispersion and dominate the density of states. The FS displays three nesting vectors $Q_1, Q_2$ and $Q_3$. b) The FS may be split into `corner' regions A,B and C and `edge' regions $\pm1, \pm2, \pm3$. \label{fig: FS}}
\end{figure}
\begin{eqnarray}
\frac{dg_1}{d\xi} &=& 2 d_1 g_1 (g_2-g_1) \nonumber\\
\frac{dg_2}{d\xi} &=& d_1(g_2^2 + g_3^2)\nonumber\\
\frac{dg_3}{d\xi} &=& - d_0 \left((n-2)  g_3^2 + 2 g_3g_4\right) + 2 d_1g_3(2g_2-g_1) \nonumber\\
\frac{dg_4}{d\xi} &=& - d_0 \left((n-1)g_3^2 + g_4^2 \right)
\label{eq: beta functions}
\end{eqnarray}
Here $d_0$ and the $d_1$ are functions of $\xi$ and we have defined
\begin{widetext}
\begin{equation}
 d_0(\xi) = \frac{d \Pi^{corner}_{pp}(0)}{d\xi} = \frac{\xi+\xi_0}{2} \quad d_1(\xi) = \frac{d \Pi^{corner}_{ph}(Q)}{d\xi} = \frac{\xi+\xi_0}{2}
\end{equation}
where perfect nesting is assumed. Imperfect nesting may be dealt with by taking $d_1<d_0$, as discussed in \cite{ChiralSC}. We have also defined $\xi_0 = \frac12 \nu_0 \ln W/\Lambda$, and we assume that the UV cutoff $\Lambda$ (and hence $\xi_0$) is the same for $d_1$ and $d_0$.

 Meanwhile, we find that for the edge-edge couplings,

\[
\frac{dh_{1A(B)}}{d\xi}=
O(\xi^0)
\qquad\frac{dh_{1C}}{d\xi}=-2v_{3A}^{2} d_0-v_{3B}^{2}d_0-2v_{4}^{2}d_1+
O(\xi^0)
\qquad\frac{dh_{2A(B)}}{d\xi}=O(\xi^0)\qquad\frac{dh_{2C}}{d\xi}=-2v_{3A}^{2}d_{0}-v_{3B}^{2}d_{0}+v_{4}^{2}d_{1}+O(\xi^0)
\]

\[
\frac{dh_{3}}{d\xi}=v_{5}^{2}d_1-2v_{4}^{2}d_1 + O(\xi^0) \qquad\frac{dh_{4}}{d\xi}=O(\xi^0)\qquad\frac{dh_{5}}{d\xi}=-v_{3}^{2}d_0-2v_{3A}v_{3B}d_0 + O(\xi^0)\qquad\frac{dh_{6}}{d\xi}=-2v_{3A}v_{3B}d_0 + O(\xi^0).
\]
We emphasize that there are log square divergences in the beta functions coming from corner-corner loops, because of which $d_1$ and $d_0$ grow with $\xi$ as $d_{0,1} \sim \xi$. This allows us to asymptotically neglect corrections to the $\beta$ functions which do not grow with $\xi$ (denoted above as $O(\xi^0)$ corrections).

 Finally, for the corner-edge couplings

\[
\frac{dv_{1A(B)}}{d\xi}=O(\xi^{0})\qquad\frac{dv_{2A(B)}}{d\xi}=O(\xi^{0})\qquad\frac{dv_{3A}}{d\xi}=-2v_{3A}(g_{4}+g_{3})d_0-2v_{3B}g_{3}d_0 + O(\xi^0)
\]

\begin{equation}
\frac{dv_{3B}}{d\xi}=-2v_{3B}g_{4}d_0-4v_{3A}g_{3}d_0 +O(\xi^0) \qquad
\frac{dv_{4}}{d\xi}=-4v_{4}g_{3}d_1 + O(\xi^0) \qquad\frac{dv_{5}}{d\xi}=2v_{5}g_{2}d_1+2v_{4}g_{1}d_1 + O(\xi^0).
\label{eq: corner edge}
\end{equation}

We can now solve the system sequentially. First we solve the corner system in a leading log approximation. The corner-corner couplings diverge along a fixed trajectory with $g_{1,2,3}\rightarrow\infty,$$g_{4}\rightarrow-\infty$
and $|g_{4}|>g_{3}>g_{2}\gg g_{1.}$ We can now determine what happens
in the corner-edge coupling space. It is convenient to define $v_{+}=2v_{3A}+ v_{3B}$ and $v_- = v_{3A}-v_{3B}$.
We find that $v_{5}$ and $v_{-}$ are relevant and flow to $+\infty$
(with $v_{-}>v_{5}$) whereas the rest of the corner-edge coupling
sector is irrelevant. Feeding this into the edge sector, we see that
\[
\frac{dh_{1C}}{d\xi}=\frac{dh_{2c}}{d\xi}=-\frac{2}{3}v_{-}^{2}d_0;\qquad\frac{dh_{3}}{d\xi}=v_{5}^{2}d_{1}\qquad\frac{dh_{5}}{d\xi}=\frac{1}{3}v_{-}^{2}d_0;\qquad\frac{dh_{6}}{d\xi}=\frac{4}{9}v_{-}^{2}d_0.
\]

\end{widetext}
with the rest of the edge-edge sector being irrelevant. Thus we see
that $h_{3,5,6}\rightarrow\infty,$ $h_{1C,2C}\rightarrow-\infty$
and all other edge-edge couplings flow to zero.

We now wish to calculate the susceptibilities for the full FS. In
\cite{ChiralSC} we already analysed the susceptibilities of the corners.
We now calculate the susceptibilities towards various ordering types
for the FS edges. We illustrate the procedure by calculating the susceptibilities
towards superconductivity. We introduce test vertices corresponding
to Cooper pairing at zero crystal momentum. There are six vertices
corresponding to six inequivalent ways of making Cooper pairs $\Delta_{1-1},\Delta_{-11},\Delta_{2-2},\Delta_{-22,}\Delta_{3-3,}\Delta_{-33},$where
it follows from Hermiticity that $\Delta_{1-1}=\Delta_{-11}^{*}$.
We now calculate the renormalization of these six test vertices under
RG. This is governed by a martix equation $\frac{d\Delta}{d\xi}=H\Delta,$where
$\Delta$ is a six component vector of test vertices and the matrix
$H$ is

\begin{equation}
H = - \left(\begin{array}{cccccc}
h_{2c} & h_5& h_6&  h_{1C}& h_{6} & h_{5}\\
h_{5}& h_{2C} & h_5 & h_{6} & h_{1C} & h_{6}\\
h_6 & h_5& h_{2C}& h_{5} & h_6&h_{1c}\\
h_{1C}  &h_6 & h_5 & h_{2C} & h_5 & h_6\\
h_6 & h_{1C} & h_6 & h_5 & h_{2C} & h_5\\
h_5 & h_6 & h_{1C} & h_6 & h_5 & h_{2C}
\end{array}\right)
\end{equation}

The most positive eigenvalue (corresponding to the strongest instability), given the asymptotic values of the $h$ fields, happens for the eigenvectors $(-1,0,1,-1,0,1)$ and $(1,-2,1,1,-2,1)$, which correspond to the two d-wave channels $d_{xy}$ and $d_{x^2-y^2}$. The eigenvalue $2(h_{1C} + h_2C - h_5 - h_6)$ is the exponent with which the d-wave superconducting instability diverges.

An analogous analysis for SDW yields an exponent of $2(h_{1C}+h_{2C})$. This is an exponent smaller than the corresponding exponent for $d$ wave superconductivity. Thus, SDW is subleading to d-wave superconductivity even on the edges.

We thus obtain the pleasing result that when d-wave superconductivity is the leading instability at the FS corners, it is also the leading instability at the edges. Put another way, the corners `pull' the edges along into the superconducting phase.

\subsection{RG away from weak coupling}

We now extend the RG away from weak coupling. When deriving the RG equations away from weak coupling, one must keep
track of $\ln$ and $\ln^{2}$ divergent terms. In the weak coupling
limit, the $\ln^{2}$ divergences are parametrically stronger, and
control the flow. However, when interactions are of moderate strength,
the phase transition can set in when $\ln$ is of order one, at which
point there is no distinction between $\ln$ and $\ln^{2}$ terms.
Therefore, to determine the phase structure away from weak coupling,
we must keep track of $\ln$ and $\ln^{2}$ divergences.

As we have discussed, while $\ln^2$ divergences come from corner-corner loops only, $\ln^1$ divergences arise also from edge-edge loops. Corner-edge loops are parametrically supressed by $\Lambda/t\ll1$, and may be ignored. We present the full RG away from weak coupling in two steps. First we analyse what happens if we ignore edge fermions altogether. Then we put the edge fermions back in.

\subsection{Patch RG away from weak coupling}
The RG equations for the corner-corner sector are obtained by extending the approach developed
for the square lattice problem \cite{Furukawa} to the number of patches
$n>2$. The number of patches matters only in diagrams with zero net
momentum in fermion loops, since it is only there that we get summation
over fermion flavors inside the loop. In this manner, we obtain the
beta functions

\begin{widetext}
\begin{eqnarray}
\frac{dg_{1}}{d\xi} & = & 2d_{1}g_{1}(g_{2}-g_{1})+d_{2}g_{1}\big[2g_{4}+(n-2)g_{1}\big]-2d_{3}g_{1}g_{2}\nonumber \\
\frac{dg_{2}}{d\xi} & = & d_{1}(g_{2}^{2}+g_{3}^{2})+2d_{2}\big[g_{1}g_{4}+(n-2)(g_{1}g_{2}-g_{2}^{2})-g_{2}g_{4}\big]-d_{3}(g_{1}^{2}+g_{2}^{2})\nonumber \\
\frac{dg_{3}}{d\xi} & = & -(n-2)d_{0}g_{3}^{2}-2d_{0}g_{3}g_{4}+2d_{1}g_{3}(2g_{2}-g_{1}),\nonumber \\
\frac{dg_{4}}{d\xi} & = & -(n-1)d_{0}g_{3}^{2}-d_{0}g_{4}^{2}+d_{2}\big[(n-1)g_{1}^{2}+2(n-1)(g_{1}g_{2}-g_{2}^{2})+g_{4}^{2}\big].\label{eq: beta functions_1}
\end{eqnarray}
 \end{widetext}
 Following the notations first introduced in \cite{Furukawa}
for RG studies of the square lattice, we have defined
\begin{eqnarray}
d_{0} & = & \frac{\partial\Pi_{pp}(Q)}{\partial\xi}=\frac{\xi_{0}+\xi}{2};\qquad d_{0}=\frac{\partial\Pi_{ph}(Q)}{\partial\xi}\approx\frac{\xi_{0}+\xi}{2,}\nonumber \\
d_{2} & = & \frac{\partial\Pi_{ph}(0)}{\partial\xi}=1,\qquad d_{3}=\frac{\partial\Pi_{pp}(Q)}{\partial\xi}\le1
\end{eqnarray}
 where $\xi_{0}=\frac{1}{2}\nu_0\ln\frac{\Lambda_{0}}{\Lambda}$
and we have assumed perfect nesting. We have allowed for the UV cutoff
of our theory, $\Lambda$, being different from the scale at which
the dispersion changes, $\Lambda_{0}$.

There are two qualitatively different regimes. When the bare interactions
are weak at the UV scale,
the RG does not flow to strong coupling
until a very large scale $\xi_{c}$, such that $d_{2}(\xi_{c})\ll1$,
$d_{3}(\xi_{c})\ll1$. In this limit, the $\ln^{1}$ terms in the
$\beta$ functions can be neglected, and the phase structure is controlled
by the $\ln^{2}$ divergent terms i.e. we can set $d_{2}=0,d_{3}=0$.
The system of RG equations collapses onto the system studied in \cite{ChiralSC},
and the only possible phase that can result for the corner fermions is d-wave superconductivity.

A qualitatively different behavior is obtained when interactions are
stronger. When the flow to strong coupling occurs for $\ln\sim O(1)$,
there is no difference between $\ln$ and $\ln^{2}$ terms, and both
must be taken into account simultaneously. To understand the behavior
in this regime, we set $d_{0}=d_{1}=d_{2}=d_{3}=1$. This places $\ln^{2}$
and $\ln$ divergent terms on an equal footing. Defining $g'_{i}=dg_{i}/d\xi$,
we obtain the RG equations (for $n=3$)

\begin{eqnarray}
g'_{1} & = & g_{1}(2g_{4}-g_{1})\nonumber \\
g'_{2} & = & g_{3}^{2}-g_{1}^{2}+2(g_{2}+g_{4})(g_{1}-g_{2})\nonumber \\
g'_{3} & = & g_{3}(4g_{2}-2g_{1}-2g_{4}-g_{3})\nonumber \\
g'_{4} & = & 2g_{1}^{2}-2g_{3}^{2}+4g_{2}(g_{1}-g_{2})\label{eq: simpler eqns}
\end{eqnarray}

This has finite coupling fixed points along the line $g_{1}=g_{2}=g_{3}=0$
and also along the line $g_{1}=g_{2}=g_{3}=2g_{4}$. To investigate
the stability of these finite coupling solutions, we consider small
deviations from the fixed line, $\delta g_{i}$, where $i=1,2,3,4$.
The flow of these small deviations is governed by $\delta g'_{i}=K_{ij}\delta g_{j}$,
where the matrix $K$ is given by
\begin{equation}
K=2\left[\begin{array}{cccc}
-g_{1}+g_{4} & 0 & 0 & g_{1}\\
g_{2}+g_{4}-g_{1} & g_{1}-2g_{2}-g_{4} & g_{3} & g_{1}-g_{2}\\
-g_{3} & 2g_{3} & -g_{3}+2g_{2}-g_{1}-g_{4} & -g_{3}\\
2g_{1}+2g_{2} & g_{1}-4g_{2} & -2g_{3} & 0
\end{array}\right]\label{eq: stability matrix}
\end{equation}
 The fixed line is stable only if all the eigenvalues of $K$ are
negative (or zero). However, $K$ has at least one positive eigenvalue
for both $g_{1}=g_{2}=g_{3}$ and $g_{1}=g_{2}=g_{3}=2g_{4}$, therefore
both fixed lines are unstable. The only stable fixed points of the
flow are at infinity.

To determine the possible fixed trajectories, we substitute into Eq.\ref{eq: simpler eqns}
the scaling ansatz
\begin{equation}
g_{i}(\xi)=\frac{G_{i}}{\xi_{c}-\xi}
\end{equation}
 where $\xi_{c}$ is the RG time at which the couplings diverge. This
turns the four coupled differential equations Eq.\ref{eq: simpler eqns}
into four coupled algebraic equations, which can be solved on mathematica.
The equations are
\begin{eqnarray}
G_{1} & = & G_{1}(2G_{4}-G_{1})\nonumber \\
G_{2} & = & G_{3}^{2}-G_{1}^{2}+2(G_{2}+G_{4})(G_{1}-G_{2})\nonumber \\
G_{3} & = & G_{3}(4G_{2}-2G_{1}-2G_{4}-G_{3})\nonumber \\
G_{4} & = & 2G_{1}^{2}-2G_{3}^{2}+4G_{2}(G_{1}-G_{2})\label{eq: scaling eqns}
\end{eqnarray}
 The scaling equations (\ref{eq: scaling eqns}) are solved most simply
by $G_{1}=G_{2}=G_{3}=G_{4}=0$ - however, this non-interacting fixed
point lies on the two fixed lines that were investigated above, and
which were found to be unstable. Therefore, we can neglect this trivial
solution.

We find using mathematica that there are 13 distinct nontrivial solutions
of (\ref{eq: scaling eqns}), corresponding to 13 different limiting
trajectories. However, not all of them can be accessed starting from
repulsive interactions. From Eq.\ref{eq: simpler eqns}, we note that
the $\beta$ functions for $g_{1}$ and $g_{3}$ vanish when the respective
couplings go to zero. Thus, these couplings cannot change sign, and
we can exclude any solution corresponding to negative values for $g_{1}$
or $g_{3}$. Similarly,$2g_{2}'+g_{4}'\propto2g_{2}+g_{4}$, therefore
the combination of couplings $2g_{2}+g_{4}$ also cannot change sign,
and we can exclude any solution with $2g_{2}+g_{4}<0$. These conditions
allow us to eliminate nine of the thirteen solutions to leave us with
only 4 fixed trajectories that can be accessed starting from repulsive
interactions. The four solutions correspond to values $(G_{1},G_{2},G_{3},G_{4})=(0,\frac{1}{2},0,-1),(0,\frac{1}{6},\frac{1}{3},-\frac{1}{3}),(\frac{1}{2},\frac{1}{4},0,\frac{3}{4})$
or $(3,-1,0,2)$.

We further want to determine which of these fixed trajectories are
stable. We therefore consider small deviations from the fixed trajectory,
\begin{equation}
g(\xi)=\frac{G_{i}}{\xi_{c}-\xi}+\delta g_{i}
\end{equation}
Substituting into Eq.\ref{eq: simpler eqns} and linearizing, we obtain
a flow equation of the form $(\xi_{c}-\xi)\delta g'_{i}=K_{ij}\delta g_{j}$,
where $K$ is given by Eq.\ref{eq: stability matrix} with $g_{i}\rightarrow G_{i}$.

Again, we want to use the matrix $K$ to check the stability of the
fixed trajectories. However, the matrix K will now certainly have
at least one eigenvector with positive eigenvalue - the eigenvector
corresponding to flow along the fixed trajectory. We want to exclude
this direction from our stability analysis. Therefore, we project
onto the subspace orthogonal to the fixed trajectory by multiplying
by the projection matrix $P=1-(G_{1},G_{2},G_{3},G_{4})\otimes(G_{1},G_{2},G_{3},G_{4})^{T}$.
We then examine the eigenvalues of the matrix $KP$. If this matrix
has any positive eigenvalues, then the fixed trajectory is unstable.
This stability analysis reveals that there are only two stable fixed
trajectories that can be accessed starting from repulsive interactions,
corresponding to two possible phases. These fixed trajectories have
critical couplings $(G_{1},G_{2},G_{3},G_{4})=(0,\frac{1}{6},\frac{1}{3},-\frac{1}{3})$
and $(\frac{1}{2},\frac{1}{4},0,\frac{3}{4})$ respectively. From
a numerical solution of the differential equations Eq.\ref{eq: simpler eqns},
we have confirmed that the RG does indeed converge to one of these
two fixed trajectories.

It now remains to determine what the leading instability is along
each fixed trajectory. We consider pairing in all channels, $\ln$
divergent as well as $\ln^{2}$ divergent (since we have set $d_{1}=d_{2}=d_{3}=d_{4}=1$,
there is no difference between these ordering channels). We now consider
the susceptibility towards developing a non-zero expectation value for every possible fermion bilinear i.e. particle-particle pairing and
particle hole pairing, at momentum transfer zero or momentum transfer
$Q$, in the spin singlet or spin triplet channel, and with any possible
structure in patch space. The various susceptibilities are tabulated
in table \ref{tab: susceptibilities}.

\begin{table*}
\begin{tabular}{lllll}
Ordering Channel  & Verbal Description  & Susceptibility exponent  & ($0,\frac{1}{6},\frac{1}{3},-\frac{1}{3})$ & ($\frac{1}{2},\frac{1}{4},0,\frac{3}{4}$)\tabularnewline
\hline
$\langle c_{1\uparrow}c_{1\downarrow}\rangle=\Delta_{1}=\Delta_{2}=\Delta_{3}$  & s wave SC  & $2(2G_{3}+G_{4})$  & $\frac{2}{3}$ & $\frac{3}{2}$\tabularnewline
\hline
$\langle c_{1\uparrow}c_{1\downarrow}\rangle=\Delta_{1}=-\Delta_{3},$$\Delta_{2}=0$  & d wave SC ( doubly degenerate)  & $2(G_{4}-G_{3})$  & $-\frac{4}{3}$  & $\frac{3}{2}$ \tabularnewline
\hline
$\langle c_{1\uparrow}c_{2\downarrow}\rangle$ & finite momentum pairing  & $2(G_{2}-G_{1})$  & $\frac{1}{3}$  & $-\frac{1}{2}$\tabularnewline
\hline
$\langle c_{1\uparrow}^{\dag}c_{2\downarrow}\rangle$ & SDW  & $-2(G_{2}+G_{3})$ & $-1$  & $-\frac{1}{2}$ \tabularnewline
\hline
$\langle c_{1\uparrow}^{\dag}c_{2\uparrow}\rangle$  & CDW  & $(2G_{1}-G_{2}+G_{3})$  & $\frac{1}{6}$  & $\frac{3}{4}$ \tabularnewline
\hline
$\langle c_{1\uparrow}^{\dag}c_{1\downarrow}\rangle=\Delta_{1}=\Delta_{2}=\Delta_{3}$ & s wave ferromagnet  & $-2(G_{4}+2G_{1})$ & $\frac{2}{3}$  & $-\frac{7}{2}$ \tabularnewline
\hline
$\langle c_{1\uparrow}^{\dag}c_{1\downarrow}\rangle=\Delta_{1}=-\Delta_{3},$$\Delta_{2}=0$  & d wave ferromagnet (doubly degenerate)  & $-2(G_{4}-G_{1})$  & $\frac{2}{3}$ & $-\frac{1}{2}$ \tabularnewline
\hline
$\langle c_{1\uparrow}^{\dag}c_{1\uparrow}\rangle=\Delta_{1}=\Delta_{2}=\Delta_{3}$  & charge compressibility  & $4G_{2}-2G_{1}+G_{4}$  & $\frac{1}{3}$  & $\frac{3}{4}$ \tabularnewline
\hline
$\langle c_{1\uparrow}^{\dag}c_{1\uparrow}\rangle=\Delta_{1}=-\Delta_{3},$$\Delta_{2}=0$  & d wave Pomeranchuk (doubly degenerate)  & $G_{1}+G_{4}-2G_{2}$  & $-\frac{2}{3}$  & $\frac{3}{4}$ 
\tabularnewline
\end{tabular}\caption[onecolumn]{Susceptibilities to various kinds of order scale as $(\xi_{c}-\xi)^{\alpha}$,
with $\alpha<0$ indicating an instability. We present here the susceptibilities
to various types of order at each fixed point, calculated for $d_{1}=d_{2}=d_{3}=1$
i.e. no difference between $\ln$ and $\ln^{2}$ divergent ordering
channels. Here $c^{\dag}$ is an electron creation operator, with
a number subscript $1,2,3$ indicating which patch the electron is
created on, and a subscript arrow labeling the spin state. \label{tab: susceptibilities}}
\end{table*}

We observe that the two fixed points show very different
behavior. At the first fixed point, there are instabilities in the
(doubly degenerate) d-wave pairing channel, in the SDW channel, and
in the d-wave Pomeranchuk channel, with the superconducting instability
leading. Thus, this fixed point is adiabatically connected to the
weak coupling fixed point, which also exhibits a leading instability
to (doubly degenerate) d wave superconductivity.

Meanwhile, at the second fixed point, there are instabilities in the
finite momentum pairing channel and in the SDW channel, and also in
the ferromagnet channel, with both s and d wave symmetry. The leading
instability is in the isotropic (s-wave) ferromagnet channel. The
resulting state breaks spin rotation symmetry, but preserves lattice
rotation symmetry and translation symmetry. Such a ferromagnet state
is natural to expect in the vicinity of a Van Hove singularity in
the DOS. However, it is only accessible for not too weak bare couplings. In the limit of zero coupling strength, the leading instability is the $d+id$ superconductor.

\subsection{Full FS RG away from weak coupling}

Thus far we concentrated solely on the corner fermions. We now put the edge fermions back in. We assume in the interests of analytic tractability that the edge fermions are in the same phase as the corner fermions (we demonstrated this to be the case at weak coupling). The problem then simplifies to determining the leading instability of the corner fermions, taking into account the renormalisations arising from the edge fermions. Assuming that $d_0=d_1=d_2=d_3=c=1$, so that all divergent diagrams in all channels are treated on an equal footing, and corner-corner loops and edge-edge loops are treated on the same footing also, we obtain $\beta$ functions for the corner-corner sector which take the form

\begin{eqnarray}
g'_{1} & = & g_{1}(2g_{4}-g_{1}) - 2 v_5^2\nonumber \\
g'_{2} & = & g_{3}^{2}-g_{1}^{2}+2(g_{2}+g_{4})(g_{1}-g_{2})+v_4^2\nonumber \\
g'_{3} & = & g_{3}(4g_{2}-2g_{1}-2g_{4}-g_{3}) - 2 v_{3A}^2 - 4 v_{3B}^2 + v_4^2 - 2 v_5^2\nonumber \\
g'_{4} & = & 2g_{1}^{2}-2g_{3}^{2}+4g_{2}(g_{1}-g_{2}) - 4 v_{3A}^2 - 2 v_{3B}^2\label{eq: complicated eqns}
\end{eqnarray}

We make one final approximation. Namely, in the $\beta$ functions for the corner-edge couplings, we retain only $\ln^{2}$ divergent terms arising from corner-corner loops, and neglect terms of $O(\ln^{1}).$ The RG equations for the corner-edge couplings are then given by Eq.\ref{eq: corner edge}. Again, we solve the system sequentially. First, we solve for the $g$ couplings. Then we solve for the $v$ couplings. Finally, we determine the back-action of the v-couplings on the g-couplings.

We note that the couplings $v_4$ and $2v_{3A}+v_{3B}$ are always irrelevant. The coupling $v_5 \rightarrow \infty$ is the only relevant corner-edge coupling along the ferromagnetic trajectory, whereas the couplings $v_5$ and $v_- = v_{3A}-v_{3B}$ are both relevant along the d-wave superconducting trajectory. Along the ferromagnetic trajectory, neglecting the irrelevant couplings, we obtain

\begin{eqnarray}
g'_{1} & \approx & g_{1}(2g_{4}-g_{1}) - 2 v_5^2\nonumber \\
g'_{2} & \approx & g_{3}^{2}-g_{1}^{2}+2(g_{2}+g_{4})(g_{1}-g_{2})\nonumber \\
g'_{3} & \approx & g_{3}(4g_{2}-2g_{1}-2g_{4}-g_{3}) - 2 v_5^2\nonumber \\
g'_{4} & \approx & 2g_{1}^{2}-2g_{3}^{2}+4g_{2}(g_{1}-g_{2})
\end{eqnarray}
Thus, along the ferromagnetic trajectory, the corner-edge couplings affect only the $g_1$ and $g_3$ couplings. They tend to suppress the $g_1$ coupling, thereby suppressing ferromagnetism, and they can also drive $g_3$ negative. If $g_3$ changes sign, then a number of other phases, such as s-wave superconductivity and charge density waves, become possible. Thus, the effect of edge fermions can be quite marked along the ferromagnetic sector, where they can alter the dominant phase from ferromagnetism to something else, like CDW or s-wave superconductivity. However, this will only happen if the effect of edge fermions is sufficiently strong, which requires strong couplings.

Meanwhile, along the SCd trajectory, both $v_5$ and $v_- = v_{3A}-v_{3B}$ are relevant. The resulting equations, neglecting irrelevant corrections, are
\begin{eqnarray}
g'_{1} & = & g_{1}(2g_{4}-g_{1}) - 2 v_5^2\nonumber \\
g'_{2} & = & g_{3}^{2}-g_{1}^{2}+2(g_{2}+g_{4})(g_{1}-g_{2})\nonumber \\
g'_{3} & = & g_{3}(4g_{2}-2g_{1}-2g_{4}-g_{3}) - 2 v_-^2 - 2 v_5^2\nonumber \\
g'_{4} & = & 2g_{1}^{2}-2g_{3}^{2}+4g_{2}(g_{1}-g_{2}) - \frac43 v_-^2\label{eq: complicated eqns_1}
\end{eqnarray}
Thus, the edge fermions supress $g_3$, which supresses SCd and SDW equally, but they also make $g_4$ more negative, which strengthens superconductivity. Thus, the effect of edge fermions along the SCd trajectory is to strengthen superconductivity with respect to SDW. Of course, if couplings are sufficiently strong that the edge fermions trigger a sign change in $g_1$, then an entirely new phase could arise. A CDW phase would be the most likely candidate given a negative $g_1$.

Thus we see that along the edge fermions can destabilise the ferromagnetic trajectory towards other phases, like s-wave superconductivity and CDW. Meanwhile, along the SCd trajectory, edge fermions strengthen superconductivity with respect to SDW. However, they can also induce formation of a different kind of phase, such as a CDW.

\subsection{Summary of the system behavior at the Van Hove point}

Thus, we can make the following conclusions.

\begin{itemize}

\item
In the weak coupling limit, the full FS RG reproduces exactly the patch RG for the saddle points presented in \cite{ChiralSC}. The entire FS enters the same phase as the FS corners, which is a d-wave superconductor.
\item
 Once we move away from weak coupling, the RG approach is no longer controlled, and the neglect of higher loop diagrams can no longer be justified. However, the procedure can still be applied, although the results must be treated with caution.
\item
Away from weak coupling, a patch RG analysis that focuses purely on the hexagon corners reveals two distinct instabilities - one to d-wave superconductivity, and another to ferromagnetism. The SDW is never the leading instability if the RG is allowed to run indefinitely, but it may dominate if the RG is stopped at some intermediate energy scale by higher loop effects or self energy effects. The ferromagnetic phase is the principal new result from extending the patch RG to strong coupling.
\item
The effect of the edge fermions at strong coupling can also be estimated. Along the ferromagnetic trajectory, the edge fermions supress ferromagnetism, and may destabilize it in the strong coupling limit towards a different phase, like a CDW or an s-wave superconductor. Meanwhile, along the SCd trajectory, the edge fermions strengthen SCd with respect to SDW, but they may destabilise SCd at strong coupling towards another phase, such as a CDW.
\end{itemize}

 Note, however, that
  although Van Hove singularities arise rather generically (see
e.g. Ref.~\cite{twisted_graphene}),
for most of them the divergence in the density
of states either is not accompanied by Fermi surface nesting,
due to, e.g.,  the
presence of longer-range hybridization,
or
the nesting vector is not one half of a reciprocal lattice
vector. In
 these cases,
  the physics differs quite substantially
from the discussions in this paper.
 For further details see e.g. Refs.~\cite{cobalt,Gonzalez2013,Yao}.

\section{Experimental outlook}
\label{sec:5}

 Despite its rapid development in recent years, the investigation of
unconventional superconductivity
 in systems
 on hexagonal lattices is still at a
comparably early stage~\cite{hexa-review}. A core
 challenge for hexagonal scenarios in general is to
identify whether
 superconductivity is of electronic
 origin. Because many hexagonal systems exhibit a strong
propensity to lattice distortions, phonon-mediated
$s$-wave
superconductivity
 is often
  a valid competitor.

With respect to electronically-mediated superconductivity, one
 primary direction
 is a search for a potential chiral $d$-wave pairing in
graphene doped to the Van Hove point~\cite{Gonzalez,ChiralSC,Ronny1,dhlee_fawang}.
Such doping can be reached by chemical means, as
Ref. [\onlinecite{Rotenberg}] has demonstrated.
 It will be revealing to
  conduct  low temperature transport experiments capable of
detecting superconductivity on these samples. Given the large amounts
of disorder introduced by chemical doping, however, it is not clear
that superconductivity should arise in chemically doped
graphene. Alternative doping techniques,
 such as ionic liquid gating, which introduce less disorder, have not yet succeeded in reaching Van
Hove filling.

However, as elaborated on in our paper and also in earlier publications\cite{scal_kiv_raghu,Ronny1},
even well away from the Van Hove
point, there is a possibility for triplet ($f$-wave)
superconductivity
Thus, another primary direction is
 fabrication of materials which are doped less that all the way to the Van Hove
point, but
have a smaller amount of disorder.
Low-temperature transport experiments
 on such samples are highly desirable.

    With ongoing
material research fostering the hope for further compounds to
arise, there are already several promising hexagonal systems
exhibiting unconventional superconductivity which might be describable
along the analysis laid out in this paper. An interesting
candidate material is SrPtAs~\cite{SrPtAs}, where preliminary
experimental evidence indicates a non-zero magnetic moment below $T_c$,
combined with the absence of line nodes. In particular, nuclear
resonance experiments seem to
be
 consistent
 with chiral $d$-wave superconductivity~\cite{neupert}.

Organic charge-transfer complexes such as the Bechgaard
salts~\cite{PhysRevLett.95.247001,cho}, the layered triangular superconductors
$\kappa$-(BEDT-TTF)$_2 X$~\cite{PhysRevLett.85.5420}, and the water-intercalated sodium cobaltates~\cite{Takada.nature.434.53,cobalt}
 are triangular lattice compounds, which, according to our study, should also display $d-$wave and $f-$wave superconductivity at different dopings.
While the interaction over bandwith ratio is rather high in these materials, the
insights from Kohn-Luttinger and parquet-RG considerations, favoring unconventional superconductivity, might still be valuable.

Artificial hexagonal optical
lattices loaded with fermionic isotopes of ultra-cold atomic gases
could establish another realization in nature of the
scenarios we describe in this paper. While the
challenge is to reduce the effective temperature $T/T_{\text{F}}$ to
make the Fermi surface instabilities accessible, all
other parameters are likely to be easily matched with an outset that
tends to exhibit unconventional superconductivity~\cite{PhysRevLett.111.185307}.

\section{conclusions}
\label{sec:6}
We have studied the emergence of superconductivity in hexagonal lattice systems (both honeycomb and triangular) over a wide range of doping levels. Away from Van Hove doping, the Kohn Luttinger framework provides a satisfactory formalism for investigating the emergence of superconductivity. Superconductivity arises if the attraction generated in a particular channel at second order in perturbation theory exceeds the bare repulsion in the corresponding channel. Thus, Kohn Luttinger superconductivity at generic doping levels in a hexagonal lattice system is a threshold phenomenon, and moreover depends on the details of the lattice scale interactions. We find that in a pure Hubbard model, superconductivity arises very generally, whereas including further neighbor interactions disfavors superconductivity. However, we also find that, if superconductivity does arise at a generic doping, it is likely to be in the f-wave channel. Thus, hexagonal lattice systems are expected to provide a good platform for f-wave superconductivity, which as far as we know has never yet been observed.

Meanwhile, close to Van Hove filling, the Kohn Luttinger formalism is inadequate, owing to the divergence of perturbation theory, and also because of the nesting of the Fermi surface, which strongly couples the particle-particle and particle hole channels. We have constructed and analyzed a parquet RG which provides an asymptotically exact description of the physics at weak coupling. The parquet analysis introduced in this paper includes the full Fermi surface in the calculation, unlike \cite{ChiralSC}, which concentrated on the parts of the Fermi surface close to the saddle points. The full Fermi surface parquet RG confirms the conclusions of \cite{ChiralSC}, establishing that the weak coupling physics is highly universal, with any choice of bare repulsive interactions producing an instability to doubly degenerate d-wave superconductivity. Moreover, the feedback of particle hole channels into the particle-particle sector ensures that the critical temperature is strongly enhanced over the Kohn Luttinger estimates. Meanwhile, extending the analysis away from weak coupling, we find that the main competitors to superconductivity if interactions are not that weak are ferromagnetism, or, potentially, charge density waves.

We expect the ideas laid out in this paper will be relevant for all investigations of superconductivity in hexagonal lattice systems. We note in particular that chiral d-wave superconductivity may already have been observed in SrPtAs \cite{SrPtAs}, in the vicinity of Van Hove doping. We hope that other ideas discussed above, such as f-wave pairing away from Van Hove doping, will also be observed in the not too distant future.

Note added: It was brought to our attention after the completion of this work that the emergence of f-wave triplet pairing on a triangular lattice was also discussed in \cite{Ikeda} 
in the context of superconductivity in  Na$_{0.35}$CoO$_{2}. 1.3$H$_2$O.  These authors included terms of order $U^3$ and argued that these terms increase $T_c$
 for f-wave pairing. 

We acknowledge with thanks fruitful discussions with L.S. Levitov, S. Raghu, E. Berg, S. Kivelson, and I. Mazin
 A.V.C. is supported by the DOE grant DE-FG02-ER46900, RT by DFG-SPP
 1458 and the
 European Research Commission through ERC-StG-2013-336012, and RN is supported by a PCTS fellowship.

\bibliographystyle{prsty}
\bibliography{fwave-ref}

\end{document}